\documentclass[journal]{IEEEtran} %
\usepackage{graphicx}
\usepackage{cite}
\usepackage{amsmath}
\usepackage{amssymb}
\usepackage{amsthm}
\usepackage{mathtools}	%
\usepackage[tight,footnotesize]{subfigure}
\usepackage{url}
\usepackage{bm} %
\usepackage{comment}
\let\labelindent\relax
\usepackage{enumitem}
\usepackage{tikz}
\usetikzlibrary{automata, positioning, arrows}

\usepackage{hyperref}
\hypersetup{
colorlinks=true
}

\theoremstyle{remark}
\newtheorem{remarkx}{Remark}
\newenvironment{remark}
{\pushQED{\qed}\remarkx}
{\popQED\endremarkx}

\theoremstyle{definition}
\newtheorem{defn}{Definition}
\newtheorem{assump}{Assumption}
\newtheorem*{problem*}{Problem}

\theoremstyle{plain}
\newtheorem{theorem}{Theorem}
\newtheorem{lemma}{Lemma}
\newtheorem{coroll}{Corollary}

\newcommand{\scalemath}[2]{\scalebox{#1}{\mbox{\ensuremath{\displaystyle #2}}}}
\newcommand{\dt}[1][{}]{\frac{d^{#1}}{dt^{#1}}}
\newcommand{\defeq}{:=}
\newcommand{\inv}[1]{#1^{-1}}    %

\newcommand{\transpose}[1]{#1^\top}

\newcommand{\mbr}[1][{}]{\mathbb{R}^{#1}}    %
\newcommand{\set}[1]{\mathcal{#1}}
\newcommand{\dist}{{\rm dist}}

\newcommand{\chiup}{\raisebox{2pt}{$\chi$}}
\newcommand{\cvf}{\chiup_c}                                    %
\newcommand{\rvfi}{\chiup_{\mathcal{R}_i}}             %
\newcommand{\rvf}{\chiup_{\mathcal{R}}}                 %
\newcommand{\rvfp}{\chiup_{\mathcal{R}_\delta}}                 
\newcommand{\pvf}{\chiup_\mathcal{P}}                            %
\newcommand{\nrvfi}{\hat{\chiup}_{\mathcal{R}_i}}     %
\newcommand{\nrvf}{\hat{\chiup}_{\mathcal{R}}}     %
\newcommand{\npvf}{\hat{\chiup}_\mathcal{P}}                %

\newcommand{\pss}{\mathcal{C}_\mathcal{P}}                    %
\newcommand{\rssi}{\mathcal{C}_{\mathcal{R}_i} }    %
\newcommand{\rss}{\mathcal{C}_{\mathcal{R}} }        %
\newcommand{\css}{\mathcal{C}_c}                            %
\newcommand{\rssp}{\mathcal{C}_{\mathcal{R}_\delta}} %

\newcommand{\zinbump}{\text{\large $\sqcup$}}            %
\newcommand{\zoutbump}{\text{\large $\sqcap$}}            %

\newcommand{\reactbdr}{\mathcal{R}}            %
\newcommand{\repulbdr}{\mathcal{Q}}            %
\newcommand{\reactin}{ \prescript{\rm in}{}{\mathcal{R}} }            %
\newcommand{\reactex}{ \prescript{\rm ex}{}{\mathcal{R}} }            %
\newcommand{\repulin}{ \prescript{\rm in}{}{\mathcal{Q}} }            %
\newcommand{\repulex}{\prescript{\rm ex}{}{\mathcal{Q}} }            %

\newcommand{\cmnt}[1]{}

\begin{document}
\title{Guiding Vector Fields for Following Occluded Paths}

\author{Weijia Yao, \IEEEmembership{Member,~IEEE}, Bohuan Lin, Brian D. O. Anderson, \IEEEmembership{Life Fellow,~IEEE}, Ming Cao, \IEEEmembership{Fellow,~IEEE}
\thanks{W. Yao and M. Cao are with the Institute of Engineering and Technology (ENTEG), B. Lin is with the Bernoulli Institute of Mathematics, Computer Science and Artificial Intelligence (BI), all at the University of Groningen, the Netherlands. \texttt{weijia.yao.new@outlook.com}, \texttt{\{b.lin, m.cao\}@rug.nl} B. D. O. Anderson is with the School of Engineering, Australian National University, Acton,  ACT 2601, Australia. {\tt\small brian.anderson@anu.edu.au}. W. Yao and B. Lin were supported in part by the China Scholarship Council.}%
}

\maketitle

\begin{abstract}
Accurately following a geometric desired path in a two-dimensional space is a fundamental task for many engineering systems, in particular mobile robots. When the desired path is occluded by obstacles, it is necessary and crucial to temporarily deviate from the path for obstacle/collision avoidance. In this paper, we develop a composite guiding vector field via the use of smooth bump functions, and provide theoretical guarantees that the integral curves of the vector field can follow an arbitrary sufficiently smooth desired path and avoid collision with obstacles of arbitrary shapes. These two behaviors are \emph{reactive} since path (re)-planning and global map construction are not involved. To deal with the common deadlock problem, we introduce a switching vector field, and the Zeno behavior is excluded. Simulations are conducted to support the theoretical results.
\end{abstract}

\IEEEpeerreviewmaketitle

\section{Introduction} \label{sec_intro}
Path-following algorithms are fundamental in robot motion control as they guide or control a robot to converge to and propagate along a \emph{geometric desired path}. Without requiring any temporal information on the desired path, path-following algorithms can sometimes overcome the inherent performance limitations in trajectory tracking algorithms \cite{aguiar2008performance}. They are also preferable in executing repetitive robot motion, such as circumference patrolling. Among several tested path-following algorithms in \cite{Sujit2014}, the \emph{vector-field guided path-following algorithms} are claimed to achieve low path-following error while requiring the least control effort. In these algorithms, sufficiently smooth vector fields are designed such that their integral curves converge to the desired path, and thereby the vector fields act as high-level guidance signals \cite{Y.A.2017,yao2020singularity,lawrence2008lyapunov,yao2020auto,Goncalves2010}. These algorithms are also practical, and effective in applications such as fixed-wing aircraft guidance and control \cite{de2017guidance,lawrence2008lyapunov}.

Usually, path-following algorithms are designed based on a predefined desired path taking no account of obstacles. Then notice is taken of (initially) unforeseen obstacles, with a view to designing a modification to the original algorithms; the modified algorithm then takes trajectories around the obstacle and return to the original desired path having passed the obstacle. This is normally achieved one obstacle at a time, and some extension to moving obstacles is possible \cite{tanveer2019analysis,sgorbissa2013integrated,lapierre2007combined,zhu2019evaluation,wilhelm2019vector}. Without considering the path-following requirement, there exist many obstacle/collision-avoidance algorithms in the field of motion planning, such as the Artificial Potential Field (APF) method \cite{khatib1986real}, the navigation function method \cite{Koditschek1990,rimon1992exact,paternain2017navigation} and the Dynamic Window (DW) method \cite{fox1997dynamic}. These approaches usually require a global map including obstacles to allow planning of a feasible path, and extensions of some of these approaches also allow only local information of obstacles \cite[Chapter 6]{siegwart2011introduction}. However, these collision-avoidance/path-planning algorithms alone are not automatically compatible with path-following algorithms. This is because a starting point and a destination point are required in collision-avoidance algorithms to plan a feasible path or determine possible moving directions between these two points, while they are not required in path-following algorithms for which a desired path is explicitly specified. In particular, vector-field guided path-following algorithms usually enable trajectories from \emph{almost all} starting points in the workspace to converge to and propagate along the desired path, rather than from only one predefined starting point and converging to a single destination point \cite{Y.A.2017,lawrence2008lyapunov}.

Only a few existing studies integrate path-following algorithms and collision-avoidance algorithms. An approach is proposed in \cite{tanveer2019analysis} to deform slightly the desired path such that the obstacle-avoidance behavior is realized at the cost of compromising the path-following accuracy. Only experimental studies using wheeled robots in an environment scattered with unforeseen static and moving obstacles are provided. In \cite{sgorbissa2013integrated}, the idea of locally deforming the desired path is adopted, and a force field is utilized to realize path-following and collision-avoidance functionalities simultaneously. This approach is experimentally verified to be computationally efficient. Nevertheless, since only straight lines are considered as desired paths between adjacent waypoints, this approach might be restrictive in applications, such as satellites circulating along orbits. In \cite{moe2016set}, a switching guidance system with a path-following mode and a collision-avoidance mode for an unmanned surface vessel is designed. The system is effective if the positions and velocities of obstacles are known, and the desired paths and obstacles are of typical geometric shapes, such as straight lines and circles. The two-mode switching methodology is also adopted in \cite{wiig2019collision}, where the authors develop the constant avoidance-angle reactive collision-avoidance algorithm and combine it with pure-pursuit or line-of-sight path-following algorithms. Mathematical analysis is conducted only for a sparse scenario with locally sensed circular obstacles. Another unified framework is proposed in \cite{lapierre2007combined}, where the authors combine the Deformable Virtual Zone (DVZ) method and the Lyapunov backstepping design. A heuristic switching mechanism is introduced when the path-following controller and the collision-avoidance controller generate antagonistic control commands.

Some studies focus on creating, modifying, or combining vector fields to realize both path-following and collision-avoidance behaviors. In \cite{zhu2019evaluation}, given accurate information of static obstacles, two vector fields, one for path following and the other for collision avoidance, are combined with weights determined by a decay function such that the adverse effects caused by the overlapping of two vector fields are mitigated. Different decay functions are numerically evaluated, but there is little theoretical guarantee for collision avoidance or path following. This approach is further developed in \cite{wilhelm2019vector} to ensure minimal deviation from the desired path during the obstacle-avoidance process, where the locations and sizes of static circular obstacles are known. The paper also proposes a numerical solution to detect singular points generated by the weighted sum of two vector fields. Nevertheless, only circular desired paths and circular obstacles are considered. In \cite{panagou2014motion}, a family of 2D vector fields is suitably blended to yield almost global feedback motion plans provided that the global information of the environment populated with static circular obstacles is given. This approach can steer a unicycle robot to a desired configuration.%

\emph{Contributions:}
This paper proposes a general and unified framework in the first instance in $\mathbb R^2$ using a \emph{composite guiding vector field} to enable trajectories to follow any sufficiently smooth desired paths occluded by static or moving obstacles of arbitrary shapes. The guiding vector field is suitable in practical situations where obstacles are detected by onboard sensors under Assumptions \ref{assump1bounded}, \ref{assump1}, \ref{assump2} and \ref{assump3}. Namely, a vehicle does not need to know the global information about obstacles before it follows a desired path, but rather it obtains in real time information about the existence, shapes and velocities of obstacles locally. The composite guiding vector field is obtained by \emph{smoothly} combining two vector fields, one for path following and the other for \emph{reactive} collision avoidance, via bump functions. The use of bump functions reduces the undesirable effects of integrating two vector fields (c.f. \cite{zhu2019evaluation}). To avoid trajectories getting stuck in a region, we also introduce a switching mechanism with detailed theoretical analysis. 

The main contributions of this paper develop the following advantages of the proposed approach:

1) Our approach is general and flexible. The desired path and the boundaries of the obstacles are any sufficiently smooth one-dimensional manifolds. Therefore, they are either homeomorphic to the unit circle or the real line\footnote{An example of the latter sort of obstacle could be a river or a coastline. }, and thereby the common convexity assumption is dropped (c.f., \cite{paternain2017navigation}). Moreover, the construction of the composite vector field does not involve any specific geometric relations between the robot, the desired path, and the obstacles. %

2) The collision-avoidance behavior in this approach is \emph{reactive} in the sense that obstacles are assumed to be unpredictable, and that the guiding vector field acts as a feedback control command \emph{directly} to the system without the process of path (re)-planning or creation of global maps. More specifically, we do not require the global knowledge about all obstacles in the design phase of the algorithm, while some minor assumptions on the obstacles are imposed (see Assumptions \ref{assump1bounded}, \ref{assump1}, \ref{assump2}, \ref{assump3}). In fact, local information about obstacles is sufficient: whenever a new obstacle is encountered, the composite vector field is easily updated by adding a new component (see Remark \ref{remark_localobs}) without compromising the original theoretical guarantees for the path-following and collision-avoidance behaviors.  This enables real-time autonomous robot navigation and motion control without knowing the global map. %

3) Our analysis is based on nonlinear systems theory, and there are rigorous theoretical guarantees for both path-following and collision-avoidance motions, which are often absent in the related literature as mentioned above. In particular, we prove that there is no Zeno phenomenon in our proposed switching mechanism (see Theorem \ref{thm2}), which is introduced to deal with the common deadlock situation. 

4) Our composite vector field can be naturally extended to any higher-dimensional spaces, including the 3D Euclidean space. Spaces with dimensions higher than three correspond to abstract configuration spaces, such as robot arm joint spaces. Thus, the composite vector field is, e.g., directly applicable in the low-level control of robot arms in the joint space.

In contrast to these advantages mentioned above, we also prove a general result showing a common limitation of combining two vector fields (see Lemma \ref{lem_limitation}). We regard this result as another contribution since it gives a theoretical explanation of the common phenomenon that singular points exist when two vector fields are blended, regardless of what decay functions (or bump functions) one uses to mitigate the overlapping effects. This result can be regarded as a counterpart of the well-known limitation of motion-planning algorithms based on the negative gradient of a potential/navigation function \cite{koditschek1990robot,rimon1992exact}, both issues being fundamentally topological. Note, though, that our settings and approach are fundamentally different from those based on a potential/navigation function. For instance, our composite vector field is not the (negative) gradient of any potential/navigation function, and we do not require a destination point to which trajectories converge. We also do not restrict consideration to a compact workspace (c.f. \cite{koditschek1990robot,rimon1992exact}). One consequence of these differences is that the limitation of our approach can sometimes be removed (see Remark \ref{remark_limitationcounter}), and thereby, perhaps surprisingly, global convergence of trajectories to the desired path with the collision-avoidance guarantee is possible (see Remark \ref{remark_limitationcounter}).

This paper significantly extends our conference version \cite{yao2019integrated}  by 1) providing all the proofs omitted in \cite{yao2019integrated}; 2) introducing the switching mechanism to deal with the existence of stable equilibria; 3) including extensive simulation examples. 

\emph{Paper Structure:}
The remainder of this paper is organized as follows. Section \ref{sec_probf} formulates the problem. In Section \ref{sec_cvf}, the systematic construction of the composite vector field is elaborated. Then Section \ref{sec_mainresult} presents the convergence results of the integral curves of the composite vector field. Section \ref{sec_switch} introduces the switching mechanism to deal with the deadlock problem. The simulation results are presented in Section \ref{sec_sim}, and Section \ref{sec_con} concludes the paper and indicates future work.

\section{Problem formulation} \label{sec_probf}
In this section, we introduce related notations, essential assumptions and formulate the problem.

\subsection{Preliminaries and notations}
We use $\dot{\xi}(t)$ to denote $\dt \xi(t)$ for any differentiable function $\xi$ of time $t$. The \emph{distance} between a point $p_0 \in \mathbb{R}^n$ and a non-empty set $\mathcal{S} \subseteq \mathbb{R}^n$ is $\dist(p_0, \mathcal{S})   \defeq \inf\{ \| p - p_0 \|_2 : p \in \mathcal{S} \}$, where $\|\cdot\|_2$ is the Euclidean norm. The distance between two non-empty sets $\mathcal{A}$ and $\mathcal{B}$ is defined by $\dist(\mathcal{A}, \mathcal{B}) \defeq \inf\{ \| a-b \|_2: a \in \mathcal{A}, b \in \mathcal{B} \}$. A trajectory $\xi: [0, +\infty) \to \mathbb{R}^n$ asymptotically converges to a non-empty set $\mathcal{A} \subseteq \mathbb{R}^n$ if for any $\epsilon>0$, there exists $T>0$ such that $\dist(\xi(t),\mathcal{A}) < \epsilon$ for $t > T$. A \emph{discrete set} $\set{A}$ in $\mbr[n]$ is a set consisting of only isolated points; that is, for every point $p \in \set{A}$, there exists an open neighborhood $\set{U}_p \subseteq \mbr[n]$ of $p$, such that $\set{U}_p \cap \set{A} = \{p\}$. A discrete set is at most countable (i.e., finite or countably infinite). A function $f:\Omega \subset \mbr[m] \to \mathbb{R}^n$ is bounded away from zero in $\Omega$ if there exists a real number $c>0$, such that $\| f(x) \|>c$ for all $x \in \Omega$. The notation $I$ represents an identity matrix of suitable dimensions. The words ``with respect to'' and ``without loss of generality'' are abbreviated as ``w.r.t'' and ``w.l.o.g'',  respectively.

\subsection{Desired paths}
The desired path $\set{P} \subseteq \mbr[2]$ is defined by
\begin{equation} \label{path}
	\mathcal{P} = \{ \xi \in \mathbb{R}^2 : \phi(\xi)=0 \},
\end{equation}
which is the zero-level set of the twice continuously differentiable function $\phi: \mathbb{R}^2 \to \mathbb{R}$. This description of the desired path $\set{P}$ does not require any parameterization. In addition, one can naturally assume that $\mathcal{P}$ is a one-dimensional connected submanifold in $\mbr[2]$. Thus, $\mathcal{P}$ is either homeomorphic to a circle if it is compact, or the real line $\mbr[]$ otherwise \cite[Theorem 5.27]{lee2010introduction}. One can use the value $|\phi(\xi)|$ at a point $\xi \in \mbr[2]$ to crudely approximate the distance $\dist(\xi, \set{P})$ between a point $\xi \in \mbr[2]$ and the path $\set{P}$ under a mild assumption shown later (i.e. Assumption \ref{assump_yuri}). For example, if a circular path $\set{P}$ of radius $R$ is described by $\phi(x,y) = x^2 + y^2 - R^2=0$, then, for every point $\xi \in \mbr[2]$ on the circle, we have $\dist(\xi, \set{P})=\phi(\xi)=0$. In addition, the distance between the point $\xi \in \mbr[2]$ and the desired path $\set{P}$ increases as $|\phi(\xi)|$ increases. For simplicity, we call $\phi(\xi)$ the (signed) \emph{path-following error} at a point $\xi \in \mbr[2]$.

\subsection{Obstacles, reactive areas and repulsive areas}
\begin{figure}[tb]
	\centering
	{
		\subfigure[$\set{P}$, $\reactbdr$, $\repulbdr$, and an obstacle]{
			\includegraphics[width=0.4\columnwidth]{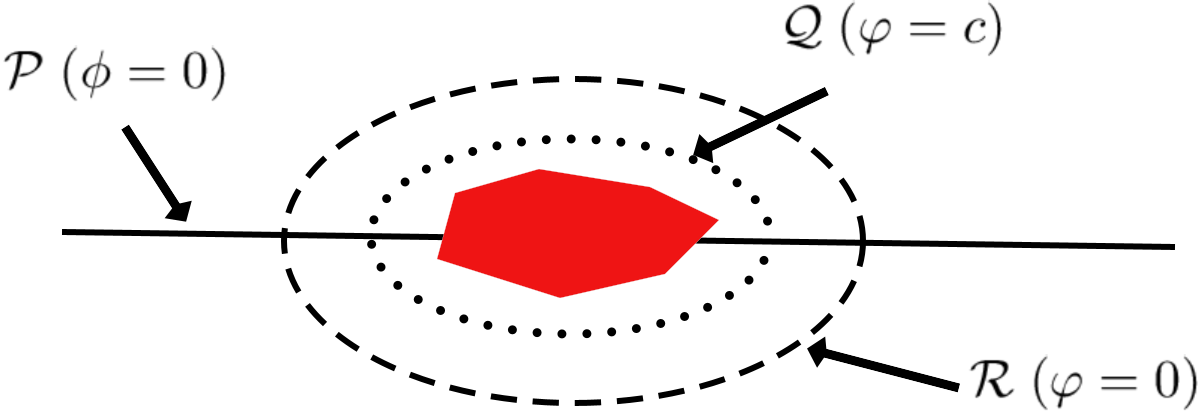}
		}%
		\subfigure[$\repulin$ and $\repulex$]{
			\includegraphics[width=0.25\columnwidth]{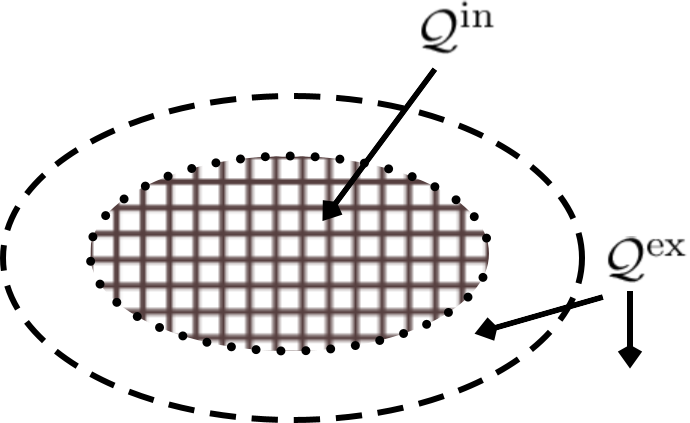}
		}%
		\subfigure[$\reactin$ and $\reactex$]{
			\includegraphics[width=0.25\columnwidth]{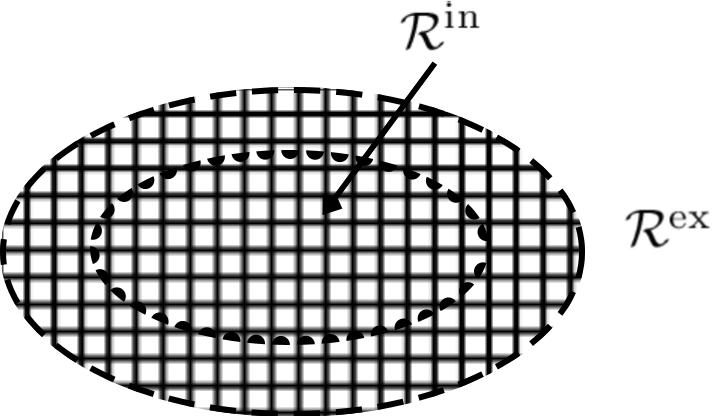}
		}%
		\caption{Illustrations of concepts.}
		\label{fig0}
	}
\end{figure}

At each time instant $t \ge 0$, we consider a finite set of obstacles $\mathcal{O}_{{\rm all}}^t = \{\mathcal{O}_i^t \subseteq \mbr[2]: i \in \mathcal{I}\}$, where $m$ is the total number of obstacles and $\mathcal{I} = \{1,2,\dots,m\}$. We assume that the obstacles are of finite sizes at every time instant\footnote{Obstacles can have infinite sizes as well; i.e., at any/some time instant $t \ge 0$, the set $\mathcal{O}_i^t$ is unbounded. This is useful, for example, when one wants to restrict a vehicle's movement within a compact space, so the obstacle is the unbounded space beyond this compact space. However, if this unbounded obstacle occludes the desired path (i.e., $\mathcal{O}_i^t \cap \set{P} \ne \emptyset$), then a trajectory might not be able to return to the desired path, and the path-following error can grow infinitely large. To avoid this undesirable consequence and for simplicity, we do not consider unbounded obstacles in this paper. }:
\begin{assump} \label{assump1bounded}
	At any time instant $t \ge 0$, the set $\mathcal{O}_i^t$ is bounded for any $i \in \mathcal{I}$. %
\end{assump}
We do not deal with the specific form of the obstacles  $\mathcal{O}_{{\rm all}}^t$, but we define some ``boundaries'' to enclose each obstacle (or to enclose a collection of obstacles if they are very close to each other) such that avoiding collision with the obstacles is simplified to avoiding collision with these boundaries regardless of the possibly complicated geometric shapes of the obstacles. Specifically, fixing $t$, we define the \textit{reactive boundary} $\reactbdr_i^t$ and the \textit{repulsive boundary} $\repulbdr_i^t$ around the obstacle $\mathcal{O}_i^t$ as follows:
\begin{subequations} \label{eq_boundaries}
\begin{align}
\reactbdr_i^t &= \{ \xi \in \mathbb{R}^2 : \varphi_i(\xi, t) = 0 \}, \label{reactive} \\
\repulbdr_i^t &= \{ \xi \in \mathbb{R}^2 : \varphi_i(\xi, t) = c_i \}, \label{repulsive}
\end{align}
\end{subequations}
where $c_i \ne 0$ is a constant and $\varphi_i:\mbr[2] \times \mbr[] \to \mbr[]$ satisfies the following assumption:
\begin{assump} \label{assump_smooth_varphi}
	The function  $\varphi_i:\mbr[2] \times \mbr[] \to \mbr[]$ in \eqref{eq_boundaries} is twice continuously differentiable.
\end{assump} 
The definitions \eqref{reactive} and \eqref{repulsive} are similar to \eqref{path} for simplicity. By Assumption \ref{assump1bounded}, $\reactbdr_i^t$ and $\repulbdr_i^t$ are compact. We also assume that at each time instant $t \ge 0$, the boundaries $\reactbdr_i^t$ and $\repulbdr_i^t$ are one-dimensional connected submanifolds in $\mbr[2]$. The reactive boundary $\reactbdr_i^t$ decomposes the plane into the ``interior'': the bounded open subset denoted by $\reactin_i^t$, and the ``exterior'': the unbounded open subset denoted by $\reactex_i^t$, and there holds $\reactbdr_i^t = \partial \reactin_i^t = \partial \reactex_i^t$ \cite[Section VI.52]{gowers2010princeton}, where $\partial (\cdot)$ denotes the boundary of a set $(\cdot)$. For convenience, we call $\reactin_i^t$ the \textit{(open) reactive area} and $\reactex_i^t$ the \textit{(open) non-reactive area}. Similarly, for the repulsive boundary $\repulbdr_i^t$, we define the \textit{(open) repulsive area} $\repulin_i^t$ and the \textit{(open) non-repulsive area} $\repulex_i^t$, and there holds $\repulbdr_i^t = \partial \repulin_i^t = \partial \repulex_i^t$ (see Fig. \ref{fig0}). We simply replace ``open'' by ``closed'' to refer to the closure of these sets (e.g. $\overline{\reactin_i^t}$ is the \textit{closed reactive area}, where $\overline{(\cdot)}$ denotes the closure of a set). Intuitively, the reactive area $\reactin_i^t$ is the area where the robot can sense the obstacles and needs to be reactive to obstacles, and the repulsive area $\repulin_i^t$ is the ``dangerous'' area where the robot is forbidden to enter.\cmnt{; in particular, if the robot's initial position is in $\repulin_i^t$, then it should eventually leave the area.} We make the following intuitively reasonable assumptions:
\begin{assump} \label{assump1}
	There holds $\mathcal{O}_i^t \subseteq \repulin_i^t \subseteq \reactin_i^t$ and $\dist(\repulbdr_i^t, \reactbdr_i^t) > 0$ for all $t \ge 0$.
\end{assump}
\begin{assump} \label{assump2}
	There holds $ \set{P} \not\subseteq \overline{\bigcup_{i \in \mathcal{I}} \reactin_i^t} $ for all $t \ge 0$.
\end{assump}
\begin{assump} \label{assump3}
	There holds $\dist(\reactin_i^t, \reactin_j^t) > 0$ for all $i \ne j \in \mathcal{I}$ and $t \ge 0$.
\end{assump}
Assumption \ref{assump1} stipulates the relative positioning of the obstacles, the reactive boundary, and the repulsive boundary. Assumption \ref{assump2} means that the desired path cannot be fully covered by obstacles; otherwise, path following is meaningless. Assumption \ref{assump3} implies that any two obstacles are sufficiently far away such that any two reactive areas are disjoint\footnote{If two reactive areas overlap, one can regard the corresponding two obstacles as one big (disconnected) obstacle and define a bigger reactive area. }.

\subsection{Problem formulation}
In the absence of obstacles, the \textbf{V}ector \textbf{F}ield guided \textbf{P}ath \textbf{F}ollowing (VF-PF) problem, as for example defined in \cite[Problem 1]{yao2020singularity}, requires one to design a vector field such that the integral curves will converge to and propagate along the desired path defined in \eqref{path}. In the presence of obstacles, the vector field needs to be modified to meet additional requirements, leading to the problem of \textbf{V}ector \textbf{F}ield guided \textbf{P}ath \textbf{F}ollowing with \textbf{C}ollision \textbf{A}voidance (VF-CAPF):
\begin{defn}[VF-CAPF] \label{def1}
	Design a continuously differentiable vector field $\chiup: \mbr \times \mbr[2] \to \mbr[2]$ for $\dot{\xi}(t) = \chiup(t, \xi(t))$ such that: 
	\begin{enumerate}[leftmargin=*]        
		\item \label{obj4} (Path-following). In the absence of obstacles, the vector field $\chiup$ solves the VF-PF problem.
		
		\item \label{obj1} (Safety). If $\xi(0) \notin \bigcup_{i \in \mathcal{I}} \overline{\repulin_i^0}$, then $\xi(t) \notin \bigcup_{i \in \mathcal{I}} \overline{\repulin_i^t}$ for $t \ge 0$.\footnote{Since the closed repulsive areas are ``dangerous'', it is naturally assumed that the initial conditions do not include these areas. Nevertheless, as shown later, our approach can still guarantee that, under some conditions, almost all trajectories starting from the closed repulsive areas will leave eventually (see Corollary \ref{cor2} and the first simulation example in Section \ref{sec_sim}).}  
		
		\item \label{obj2} (Bounded Path-following Error). There exists $M>0$ such that the path-following error satisfies $|\phi(\xi(t))| \le M $ for $t \ge 0$. Moreover, for any connected time interval $\Xi \subseteq \mbr[]$ satisfying $\xi(t) \notin \bigcup_{i \in \mathcal{I}} \overline{\reactin_i^t}$ for $t \in \Xi$, the absolute path-following error $|\phi(\xi(t))|$ is strictly decreasing over $\Xi$.
		
		\item \label{obj3} (Penetrable $\overline{\reactin_i^t}$).  Fix $i \in \mathcal{I}$, and consider trajectories $\xi(\cdot)$ starting from almost all initial conditions. If there exists $t_1 > 0$ such that $\xi(t_1) \in \overline{\reactin_i^{t_1}}$, then there exists another (possibly non-unique) time instant $t_2 > t_1$ such that $\xi(t_2) \notin \overline{\reactin_i^{t_2}}$. Additionally, the trajectory cannot cross the reactive boundary $\reactbdr_i^t$ infinitely fast; i.e., $\inf\{ t_2 - t_1 :  t_2 > t_1, \xi(t_2) \notin \overline{\reactin_i^{t_2}} \} > 0$.
			\end{enumerate}
\end{defn}

\begin{remark}
	Objective \ref{obj1} ensures that trajectories do not collide with obstacles. As obstacles might be right on the desired path, Objective \ref{obj2} requires that the path-following error should be at least bounded by some constant. In addition, the path-following error should be decreasing along a trajectory if no obstacles are nearby. Objective \ref{obj3} prevents trajectories starting from almost all initial points from staying in a reactive area forever such that the repetitive motion of following the desired path is possible (i.e., a reactive area is always penetrable). \cmnt{``Almost all'' means the set of initial points of trajectories that cannot leave the reactive area forms a zero-measure set.} Objective \ref{obj3} also excludes the Zeno behavior \cite{liberzon2003switching}.
\end{remark}
For simplicity, we consider \emph{static} obstacles first (in Sections \ref{sec_cvf}, \ref{sec_mainresult} and \ref{sec_switch}). Therefore, the superscript $t$ in the notations are removed, and the vector field $\chiup: \mbr[2] \to \mbr[2]$ becomes time-invariant. The extension to moving obstacles and the time-varying vector field will be discussed in Remarks \ref{remark_moving_obs} and \ref{remark_moving_robust}.

\section{Composite Vector Field} \label{sec_cvf}

\subsection{Path-following vector field and reactive vector field}
\cmnt{The basic task of vector-field guided path following is designing a vector field that guides a robot to move towards and circulate along the desired path. This approach has several advantages, including achieving high accuracy while requiring low control effort \cite{Sujit2014} and guaranteeing (almost) global convergence to the desired path \cite{Y.A.2017,yao2018cdc,yao2020singularity}. }
From the definitions of the desired path in \eqref{path} and the reactive boundary in \eqref{reactive}, we can define the \textit{vector fields} \cite{Y.A.2017} $\pvf, \rvfi: \mathbb{R}^2 \to \mathbb{R}^2$ associated with $\set{P}$ and $\reactbdr_i$ by
\begin{align}
\pvf(\xi) &= \gamma_0 E \nabla\phi(\xi) - k_p \phi(\xi) \nabla\phi(\xi), \label{eqpvf} \\
\rvfi(\xi) &= \gamma_i E \nabla\varphi_i(\xi) - k_{r_i} \varphi_i(\xi) \nabla\varphi_i(\xi), \; i \in \set{I} \label{eqrvf}
\end{align}
where $\gamma_i \in \{ 1,-1\}$, $i\in\{0\} \cup \set{I}$ determines the movement direction along the desired path or the reactive boundaries, $E = \left[ \begin{smallmatrix} 0 & -1  \\ 1 & 0 \end{smallmatrix} \right]$ is the $90^\circ$ rotation matrix and $k_p$, $k_{r_i}$ are positive gains. The latter term of each equation above is a signed gradient, and guides integral curves to converge towards the desired path or the reactive boundaries, whereas the first term is perpendicular to the gradient and provides a movement speed along the desired path or the reactive boundaries. \emph{For simplicity, we assume $\gamma_i=1$ for all $i \in \{0\} \cup \set{I}$ throughout the subsequent theoretical development by default}.
We call $\pvf$ the \textit{path-following vector field} and $\rvfi$ the \textit{reactive vector field}. The singular set of the vector fields $\pvf$ and $\rvfi$ are denoted by $\pss$ and $\rssi$ respectively and defined below:
\begin{align*}
\pss  &= \{ \xi \in \mbr[2] : \pvf(\xi) = 0 \} = \{ \xi \in \mbr[2] : \nabla \phi(\xi) = 0 \}, \\
\rssi &= \{ \xi \in \mbr[2] : \rvfi(\xi) = 0 \} = \{ \xi \in \mbr[2] : \nabla \varphi_i(\xi) = 0 \},
\end{align*}
where the second equation of the first line is due to the fact that $\pvf(\xi) = 0 \iff (\gamma_0 E - k_p \phi(\xi) I) \nabla\phi(\xi) = 0 \iff \left[ \begin{smallmatrix}
	-k_p \phi(\xi) & -\gamma_0 \\ \gamma_0 & - k_p \phi(\xi)
\end{smallmatrix} \right] \nabla \phi(\xi) =0 \iff \nabla\phi(\xi) = 0$, noting that the matrix $\left[ \begin{smallmatrix} 	-k_p \phi(\xi) & -\gamma_0 \\ \gamma_0 & - k_p \phi(\xi) \end{smallmatrix} \right]$ is non-singular for any $\xi \in \mathbb{R}^2$ as $\gamma_0 \ne 0$. The second equation of the second line is obtained using the same reasoning.
Each point in $\pss$ or $\rssi$ is called a \emph{singular point}, where the corresponding vector field vanishes. In this case, the singular points happen to be the \emph{critical points} of $\phi$ or $\varphi_i$, but this is not true for vector fields defined in other higher-dimensional (Euclidean) spaces \cite{yao2020auto,Goncalves2010}. Since $\set{P}$, $\reactbdr_i$ and $\repulbdr_i$ are one-dimensional connected submanifolds in $\mbr[2]$, $0$ is a regular value for $\phi$ and $\varphi_i$, and $c_i$ is another regular value for $\varphi_i$ \cite[p. 105]{lee2015introduction}. Therefore, there are no singular points on $\set{P}$, $\reactbdr_i$ and $\repulbdr_i$; i.e., $\set{P} \cap \pss = \emptyset$, $\reactbdr_i \cap \rssi = \emptyset$ and $\repulbdr_i \cap \rssi = \emptyset$.

\subsection{Behavior with a single vector field}
Let $\chiup = \pvf$ or $\chiup = \rvfi$, and we consider the following autonomous ordinary differential equation:
\begin{equation} \label{eqode_preliminary}
	\dot{\xi}(t) = \chiup(\xi(t)).
\end{equation}
Given an initial condition $\xi(0) \in \mbr[2]$, the existence (possibly up to a finite time) and uniqueness of solutions to \eqref{eqode_preliminary} is guaranteed, as $\chiup(\xi)$ is continuously differentiable (and hence locally Lipschitz continuous) w.r.t $\xi$ \cite[Theorem 3.1]{khalil2002nonlinear}. It is unclear yet whether a solution exists for $t \ge 0$, but Lemma \ref{lemdic} will show that if a solution converges to the desired path $\set{P}$, then the solution exists for $t \ge 0$. Taking $\chiup=\pvf$ as an example, it is proved in \cite{Y.A.2017,yao2020auto} that a trajectory $\xi(t)$ converges to either the desired path $\set{P}$ or the singular set $\pss$ as $t \to \infty$ under the following assumption:
\begin{assump}[\cite{Y.A.2017}] \label{assump_yuri}
	For any $\kappa>0$, there holds $\inf\{|\phi(\xi)| : \dist(\xi, \set{P}) \ge \kappa \} > 0$. Similarly, for any $\kappa>0$, there holds $\inf\{\| \nabla \phi(\xi) \| : \dist(\xi, \pss) \ge \kappa \} > 0$.
\end{assump}
Note that corresponding to the same desired path $\set{P}$, there is an infinite number of choices\footnote{For example, if $\set{P}=\phi^{-1}(0)$, then one can define $\bar{\phi}(\cdot)=\phi(\cdot) \Lambda(\cdot)$, where $\Lambda: \mbr[] \to \mbr[]$ and $\Lambda(p) \ne 0$ for any $p \notin \set{P}=\phi^{-1}(0)$. A trivial choice of $\Lambda(\cdot)$ is any non-zero constant function. Therefore, $\set{P}=\phi^{-1}(0)=\bar{\phi}^{-1}(0)$ and $\bar{\phi}$ is another function to characterize the same desired path $\set{P}$. } of the function $\phi$ in \eqref{path}. This assumption restricts one to choose a ``valid'' function $\phi$ such that when $|\phi(\xi(t))| \to 0$ as $t \to \infty$ along an infinitely-extendable trajectory $\xi(t)$, then $\dist(\xi(t), \set{P}) \to 0$ as $t \to \infty$ (guaranteed by the first part of the assumption), and when $\| \nabla \phi(\xi(t)) \| \to 0$ as $t \to \infty$ then $\dist(\xi(t), \pss) \to 0$ as $t \to \infty$ (guaranteed by the second part of the assumption).  Without this assumption, one can choose a function $\phi$ such that trajectories diverge to infinity even when the absolute path-following error $|\phi(\cdot)|$ converges to zero (see \cite[Section IV.B]{yao2018cdc}, \cite[Example 1]{yao2021dichotomy}). In addition, the assumption is not restrictive as it is satisfied for many polynomial or trigonometric functions $\phi$ \cite{yao2018cdc,Goncalves2010}; the assumption holds for all examples presented in this paper. Under this assumption, we have the following important \emph{dichotomy convergence} lemma.

\begin{lemma}[Dichotomy convergence, \cite{Y.A.2017,yao2020auto}] \label{lemdic}
	Under Assumption \ref{assump_yuri}, let $\pvf: \mbr[2] \to \mbr[2]$ be the vector field associated with the one-dimensional connected submanifold  $\mathcal{P}$ described by \eqref{path}. Given an initial condition $\xi(0) \in \mbr[2]$, any trajectory of $\dot{\xi}(t) = \pvf(\xi(t))$ converges either to $\mathcal{P}$ or the singular set $\pss$ as $t \to \infty$.
\end{lemma}
\begin{remark}[Time-invariant scaling] \label{remark2}
	Lemma \ref{lemdic} continues to hold up to a time-invariant positive scaling (e.g., the normalization) of the vector field, such that the orientation of each vector of $\pvf$ is not modified \cite[Proposition 1.14]{chicone2006ordinary}. \cmnt{Nevertheless, if the vector field is normalized, a trajectory only exists on a finite time interval when it converges to $\mathcal{C}$ \cite{Y.A.2017}.}
\end{remark}

In path following, the convergence of trajectories to the singular set $\pss$ is not desirable, and therefore, it is important to know how large the set of initial points of trajectories converging to $\pss$ is. We use the notation $\set{W}(\pss)$ to represent the set of initial conditions for trajectories of \eqref{eqode_preliminary} to asymptotically converge to the singular set $\pss$. Namely,
$
	\scalemath{0.95}{\set{W}(\pss) \defeq  \{\xi_0 \in \mbr[n]: \xi(0)=\xi_0, \dist(\xi(t), \pss) \to 0 \text{ as } t \to \infty \}.}
$
If $\pss$ consists of only one point denoted by $q$, which is also an equilibrium of \eqref{eqode_preliminary}, then we define $\set{W}(q) \defeq \set{W}(\pss)$, which is called the \emph{local inset} of $q$ \cite[p. 39]{sastry2013nonlinear}. \cmnt{If the point $q$ is hyperbolic (i.e., the eigenvalues of the linearization at this point are off the imaginary axis), then the local inset $\set{W}(q)$ is a manifold, called the \emph{stable manifold} of $q$ \cite[Theorem 7.6]{sastry2013nonlinear}.} If the singular set $\pss$ is discrete or the function $\phi$ is real analytic, a trajectory converging to $\pss$ actually converges to a single point in $\pss$, and therefore, $\set{W}(\pss) = \bigcup_{q \in \pss} \set{W}(q)$ \cite[Theorems 3 and 4]{yao2021dichotomy}. For any point $q \in \pss$, the Hessian matrix at this point is denoted by $H_{\phi}(q) \defeq \nabla^2 \phi(q)$, which is a symmetric matrix since $\phi \in C^2$. Now we state the following result regarding how ``large'' $\mathcal{W}(\pss)$ is.
\begin{lemma} \label{lemma_inset}
	Suppose the singular set $\pss$ is discrete. If the matrix $\phi(q) H_{\phi}(q)$ has at least one negative eigenvalue for every point $q \in \pss$, then $\set{W}(\pss) = \bigcup_{q \in \pss} \set{W}(q)$ is a set of measure zero. If $\phi(q) H_{\phi}(q)$ has all negative eigenvalues for every point $q \in \pss$, then $\set{W}(\pss) = \pss$.
\end{lemma}
\begin{proof}
	It is proved in \cite[Lemma 3 and Corollary 1]{Y.A.2017} that if $\phi(q) H_{\phi}(q)$ has at least one negative eigenvalue for $q \in \pss$, then $\mu(\set{W}(q))=0$, where $\mu$ is the Lebesgue measure. Since $\pss$ is discrete, it is at most countable. By the nonnegativity and subadditivity properties of the Lebesgue measure, $0 \le \mu(\set{W}(\pss)) = \mu \left( \bigcup_{q \in \pss} \set{W}(q) \right) \le \sum_{q \in \pss} \mu(\set{W}(q))=0$. Therefore, $\set{W}(\pss)$ is of measure zero. The last statement of the lemma is due to Lemma 3 and Corollary 1 in \cite{Y.A.2017}.
\end{proof}
\begin{remark}
\cmnt{If the Hessian matrix $H_{\phi}(q)$ of a critical point $q \in \pss$ is non-singular, this point $q$ is called \emph{non-degenerate}. If every critical point of $\phi$ is non-degenerate, then $\phi$ is a \emph{Morse function} \cite[Definition 1.14]{matsumoto2002introduction}.} If $\phi$ is a Morse function, then every critical point is isolated \cite[Corollary 1.12]{matsumoto2002introduction}, and thus $\pss$ is discrete. Since almost all smooth functions are Morse functions\footnote{Morse functions form an open, dense subset of the space of smooth functions \cite[Chapter 6, Theorem 1.2]{hirsch2012differential}.}, the condition of $\pss$ being discrete is not conservative. In many practical examples, $\pss$ is even finite \cite{Y.A.2017,lawrence2008lyapunov}. In particular, if $\phi$ is a Morse function and $\nabla \phi(\xi) \ne 0$ when $\| \xi \|$ is sufficiently large, then $\pss$ is finite.%
\end{remark}
Lemma \ref{lemdic} and Lemma \ref{lemma_inset} still hold if $\set{P}$, $\phi$, $H_{\phi}$,  $\pvf$, and $\pss$  are replaced with $\reactbdr_i$, $\varphi_i$, $H_{\varphi_i}$, $\rvfi$, and $\rssi$ respectively.

\subsection{Smooth zero-in and zero-out functions}

In preparation for studying the simultaneous effects of two vector fields, we introduce some special functions to ``blend'' different vector fields, which is inspired by the result below.
\begin{lemma}[{\cite[Proposition 2.25]{lee2015introduction}}] \label{lem1}
	Given a non-empty open subset $\mathcal{B} \subseteq \mbr[n]$ and a non-empty closed subset $\mathcal{A}  \subseteq \mbr[n]$ such that $\mathcal{A} \subseteq \mathcal{B}$, there exists a smooth (i.e., infinitely differentiable) function $\zoutbump: \mbr[n] \to \mbr[]$ such that $\zoutbump(x) \equiv 1$ for $x \in \mathcal{A}$, $ 0 \le \zoutbump(x) \le 1$ for $x \in \mathcal{B} \setminus \mathcal{A}$ and $\zoutbump(x) \equiv 0 $ for $x \in \mbr[n] \setminus \mathcal{B}$. 
\end{lemma}
The function $\zoutbump$ in Lemma \ref{lem1} is a \textit{smooth bump function}, which is a smooth real-valued function that attains $1$ on a compact set and attains zero beyond an \emph{open neighborhood} of that set \cite[pp. 40-47]{lee2015introduction}. It is obvious that there also exists an ``inverted'' bump function $\zinbump: \mbr[n] \to \mbr[]$ which attains $0$ on a compact set and attains $1$ beyond an \emph{open neighborhood} of that set (e.g., $\zinbump=1-\zoutbump$). These functions are ideal to ``blend'' different vector fields by keeping or removing some parts of the vector field to possibly reduce the undesirable effects of overlapping, while retaining the original smoothness properties of the vector fields. We have the following corollary.
\begin{coroll} \label{cor1}
	For any reactive boundary $\reactbdr_i$ in \eqref{reactive} and repulsive boundary $\repulbdr_i$ in \eqref{repulsive}, $i \in \mathcal{I}$, there exist smooth functions $\zinbump_{\repulbdr_i}, \zoutbump_{\reactbdr_i}: \mbr[2] \to [0, \infty) $ defined below:
	\begin{equation} \label{eq_bump}
	\scalemath{0.8}{
		\begin{split}
			 \zinbump_{\repulbdr_i}(\xi) = 
			\begin{dcases}
			0 & \xi \in \overline{\repulin_i} \\
			S_i(\xi) & \xi \in \repulex_i \cap \reactin_i, \\ %
			1	& \xi \in \overline{\reactex_i}
			\end{dcases}
			\quad
			 \zoutbump_{\reactbdr_i}(\xi) = 
			\begin{dcases}
			1			&  \xi \in \overline{\repulin_i}  \\
			Z_i(\xi)     & \xi \in \repulex_i \cap \reactin_i,  \\ %
				0             & \xi \in \overline{\reactex_i} \\
			\end{dcases}
		\end{split}
	}
	\end{equation}
where $S_i: \repulex_i \cap \reactin_i \to (0, 1) $ and $Z_i: \repulex_i \cap \reactin_i \to (0, 1)$ are smooth functions. 
\end{coroll}
Intuitively, we call $\zoutbump_{\reactbdr_i}$ a \textit{smooth zero-out function} w.r.t $\reactbdr_i$ and $\zinbump_{\repulbdr_i}$ a \textit{smooth zero-in function} w.r.t $\repulbdr_i$. Since $\zinbump_{\repulbdr_i}$ is smooth, it is evident that $S_i(\xi)$ vanishes smoothly to $0$ as $\xi$ approaches $\repulbdr_i$, and smoothly approaches value $1$ as $\xi$ approaches $\reactbdr_i$, and the converse applies for $Z_i(\xi)$.

\subsection{Composite vector field}
We use smooth zero-in and zero-out functions to ``blend'' different vector fields and obtain the composite vector field $\cvf: \mathcal{D} \subseteq \mbr[2] \to \mbr[2]$ as follows:
\begin{equation} \label{eqcompvf}
\cvf(\xi) = \left( \prod_{i \in \mathcal{I}} \zinbump_{\repulbdr_i}(\xi) \right) \npvf(\xi) + \sum_{i \in \mathcal{I}} \left( \zoutbump_{\reactbdr_i}(\xi) \nrvfi(\xi) \right) ,
\end{equation}
where $\hat{(\cdot)}$ is the normalization notation (i.e., $\hat{v} = v/ \| v \|$ for a nonzero vector $v \in \mbr[n]$), and $\mathcal{D} = \mbr[2] \setminus (\bigcup_i \rssi \bigcup \pss)$ is the domain on which the composite vector field is well-defined. Under Assumption \ref{assump3}, \eqref{eqcompvf} is equivalent to:
\begin{equation} \label{eqcompvf_detail}
\scalemath{0.9}{
\begin{split}
\cvf(\xi) \stackrel{\eqref{eq_bump}}{=} 
\begin{dcases}
\nrvfi(\xi) & \xi \in \overline{\repulin_i} \cap \mathcal{D}  \\
S_i(\xi) \npvf(\xi)  + Z_i(\xi) \nrvfi(\xi) & \xi \in \repulex_i \cap \reactin_i \cap \mathcal{D}  \\
\npvf(\xi) & \xi \in \mathcal{D} \setminus ( \bigcup_{k \in \mathcal{I}} \reactin_k ),
\end{dcases}
\end{split}
}
\end{equation}
for each $i \in \mathcal{I}$. From \eqref{eqcompvf_detail}, one observes that only the reactive vector field $\rvfi$ is active in the corresponding closed repulsive areas $\overline{\repulin_i}$, and only the path-following vector field $\pvf$ is active in the union of the closed non-reactive areas $\bigcup_{i \in \mathcal{I}} \overline{\reactex_i}$, whereas both the path-following vector field $\pvf$ and the reactive vector field $\rvf$ are active in the intersection of the reactive area and the non-repulsive area; namely, $\mathcal{M}_i \defeq\repulex_i \cap \reactin_i$, called the \textit{(open) mixed area} for convenience. The closure of this area $\overline{\mathcal{M}_i}$ is called the \textit{closed mixed area}. Note that the composite vector field $\cvf$ is not ``mixed'' on the reactive and repulsive boundaries.
We consider the following system:
\begin{equation} \label{eqode}
	\dot{\xi}(t) = \cvf(\xi(t)), \quad \xi(0) \in \mathcal{D}.
\end{equation}
We define the \emph{composite singular set} below:
\begin{equation} \label{eq_css}
	\css \defeq \{\xi \in \mathcal{D} : \cvf(\xi) = 0 \},
\end{equation} 
which contains all the equilibria of \eqref{eqode}. Note that this set may contain some singular points in $\bigcup_i \rssi \bigcup \pss$, as well as new singular points due to the blending of the two vector fields in the union of the mixed area $\bigcup_i \set{M}_i$.
\begin{remark} \label{remark_localobs}
	Only locally sensed obstacles need to be considered for the computation of the composite vector field $\cvf$.  From the compact expression in \eqref{eqcompvf}, it seems necessary to know \emph{all} the obstacles in the workspace, but this is not true if one observes the expanded form in \eqref{eqcompvf_detail}. When a trajectory enters a reactive area, meaning that a robot can detect an obstacle, the composite vector field $\cvf$ in \eqref{eqcompvf_detail} only depends on the current reactive vector field $\rvfi$ and the path-following vector field $\pvf$ but \emph{not} on the knowledge of other obstacles. 
\end{remark}

\section{Analysis of the composite vector field} \label{sec_mainresult}

\begin{figure}[tb]
	\centering
	{
		\subfigure[$\npvf$]{
			\includegraphics[width=0.3\columnwidth]{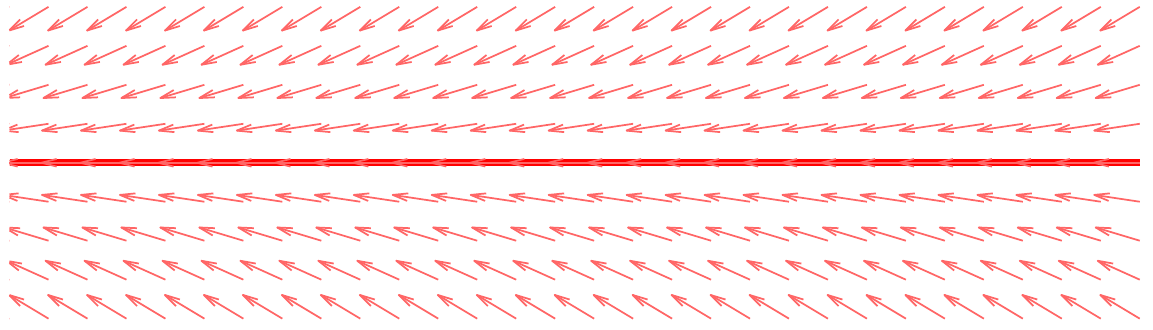}
		}%
		\subfigure[$\nrvf$]{
			\includegraphics[width=0.3\columnwidth]{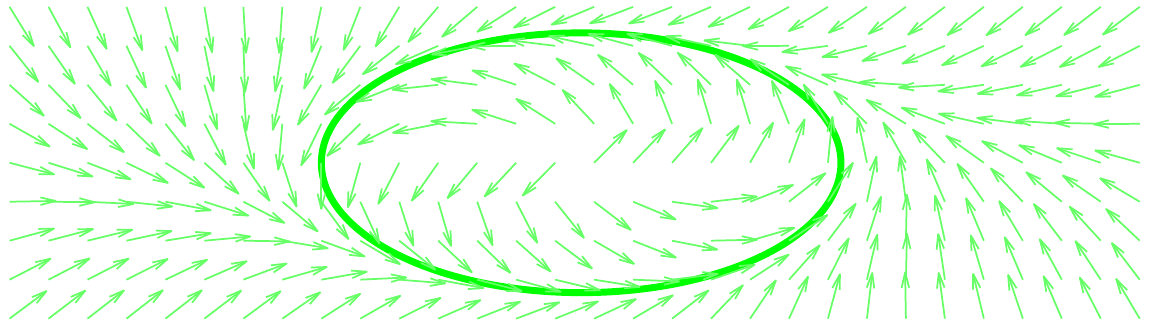}
		}%
		\subfigure[$\zinbump_{\repulbdr}(\xi) \npvf(\xi)$]{
			\includegraphics[width=0.3\columnwidth]{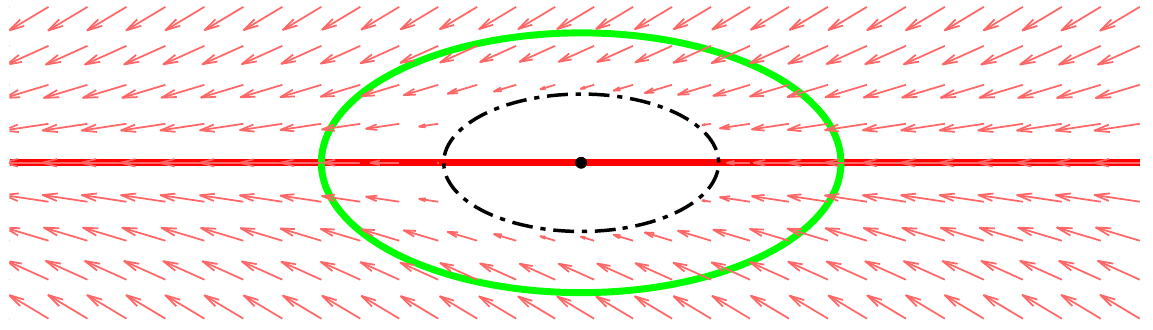}
		}\\
		\subfigure[$\zoutbump_{\reactbdr}(\xi) \nrvf(\xi)$]{
			\includegraphics[width=0.4\columnwidth]{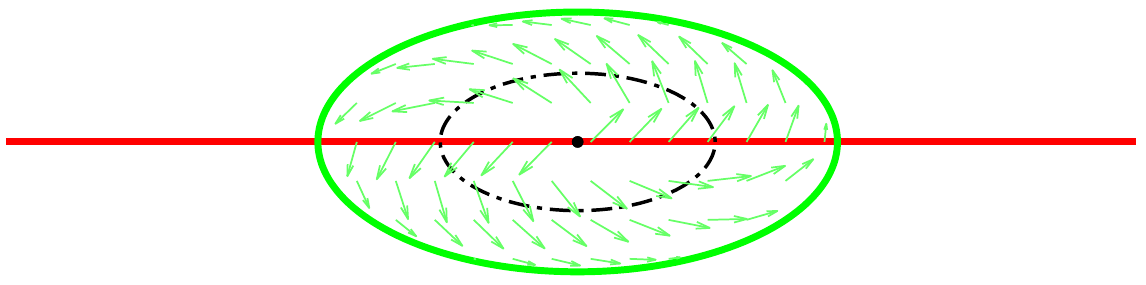}
		}
		\subfigure[$\cvf$]{
			\includegraphics[width=0.4\columnwidth]{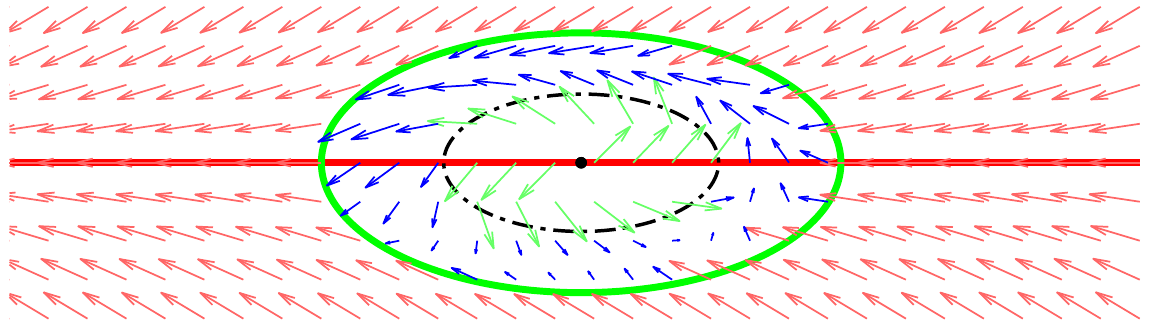}
			\label{fig:e}
		}%
	}
	\caption{Construction of the composite vector field $\cvf$ in \eqref{eqcompvf_single}. Each arrow in the subfigures represents a vector of the corresponding vector field evaluated at the position of the \emph{tail} of the arrow. In \subref{fig:e}, the green arrows belong to the reactive vector field $\rvf$ and are all in $\overline{\repulin}$. The red arrows belong to the path-following vector field $\pvf$ and are all in $\overline{\reactex}$. The blue arrows belong to the weighted sum of $\pvf$ and $\rvf$ and are all in the mixed area $\mathcal{M}=\repulex \cap \reactin$. }
	\label{fig1}
\end{figure}

Under Assumption \ref{assump3} and in view of \eqref{eqcompvf_detail}, different reactive vector fields $\rvfi$ do not overlap. Therefore, w.l.o.g, we only consider the case of one obstacle; i.e., the index set $\mathcal{I}$ is a singleton. Thus, \emph{the subscripts $i$ are removed from the notations in this section for simplicity}. We also assume that the obstacle is sufficiently close to the desired path (precisely, $\reactin \cap \set{P} \ne \emptyset$) such that collision avoidance is necessary. Now, the composite vector field \eqref{eqcompvf_detail} simplifies to
\begin{equation} \label{eqcompvf_single}
\scalemath{0.9}{
\cvf(\xi) =
\begin{dcases}
	\nrvf(\xi) & \xi \in \overline{\repulin} \cap \mathcal{D} \\
	S(\xi) \npvf(\xi) + Z(\xi) \nrvf(\xi) & \xi \in \repulex \cap \reactin \cap \mathcal{D} \\
	\npvf(\xi) & \xi \in \overline{\reactex} \cap \mathcal{D},
\end{dcases}
}
\end{equation}
where $\mathcal{D} = \mbr[2] \setminus (\rss \cup \pss)$. An intuitive illustration is shown in Fig. \ref{fig1}. The repulsiveness of a set is defined as follows.
\begin{defn}[Repulsiveness] \label{def3}
	A nonempty set $\mathcal{A} \subseteq \mbr[2]$ is \textit{repulsive} w.r.t the dynamics $\dot{\xi} = f(\xi)$, where $f: \mbr[2] \to \mbr[2]$ is Lipschitz continuous, if for each point $\xi_0 \in \mathcal{A}$, there exists $T>0$, such that the trajectory $\xi(t)$ with the initial condition $\xi(0) = \xi_0$ satisfies $\xi(t) \notin \mathcal{A}$ for $t \ge T$. %
	If this holds for almost every point in $\set{A}$ except for a set of measure zero, then $\set{A}$ is called \emph{almost repulsive}. Namely, $\set{A}$ is \emph{almost repulsive} if there exists a subset $\set{B} \subseteq \set{A}$ of measure zero such that $\set{A} \setminus \set{B}$ is repulsive.
\end{defn}

The following lemma states the positive invariance property of the non-repulsive area $\repulex$. Namely, for all $\xi(0) \in \repulex$, it follows that $\xi(t) \in \repulex$ for $t \ge 0$.
\begin{lemma}[Positive invariance of $\repulex$] \label{lem3}
	If $\mathcal{W}(\rss) \cap \repulbdr = \emptyset$, then $\repulex$ is positively invariant w.r.t \eqref{eqode}. 
\end{lemma}
\begin{proof}
	We prove this by contradiction. Suppose $\repulex$ is not positively invariant, then there exists $\xi(0) \in \repulex$ and $T>0$ such that $\xi(T) \in \repulbdr$. Note that in the closed repulsive area $\overline{\repulin}$, the differential equation is simplified to $\dot{\xi} = \nrvf(\xi)$. Also note that the trajectory will leave $\repulbdr$ later; otherwise, this implies $\repulbdr$ contains a periodic orbit and thus contradicts Lemma \ref{lemdic}. Therefore, there exists some time instant $T' > T$ such that $\xi(T') \notin \repulbdr$. Next, we prove that the trajectory cannot go into the interior $\repulin$. Suppose, on the contrary, $\xi(T') \in \repulin$. Since $\mathcal{W}(\rss) \cap \repulbdr = \emptyset$, the trajectory will not converge to $\rss$. Therefore, the trajectory will reach the boundary $\repulbdr$ again at some time instant $T''>T'$; otherwise, again there is a contradiction of Lemma \ref{lemdic}. Hence, on the time interval $\Xi \defeq [T, T'']$, $\xi(t) \in \overline{\repulin}$, and $\xi(T), \xi(T'') \in \repulbdr$. For convenience, we denote the segment of trajectory $\xi(t)$ over $\Xi$ by $\xi(\Xi) \defeq \{\xi(t) \in \mbr[2] : t \in \Xi \}$, which is the image of the time interval $\Xi$ under $\xi$. Since $\xi(\Xi)$ is closed and $\xi(\Xi) \subseteq \overline{\repulin}$, the segment of trajectory $\xi(\Xi)$ is compact in $\mbr[2]$. For what follows, we consider the trajectory over the time interval $\Xi$. Again, note that the differential equation \eqref{eqode} is simplified to $\dot{\xi} = \nrvf(\xi)$ over this time interval. Choose the Lyapunov function candidate $V(\varphi(\xi)) =  \varphi^2(\xi) / 2$, then 
	\begin{equation} \label{eqdvdvarphi}
	\dt[] V(\varphi(\xi)) = -k_r \frac{1}{\| \rvf(\xi) \| } \varphi^2(\xi) \| \nabla \varphi(\xi) \|^2.
	\end{equation}
	Due to the uniqueness of solutions and $\mathcal{W}(\rssi) \cap \repulbdr = \emptyset$, we have $\dt[] V(\varphi(\xi(t))) < 0$ for $t \in \Xi$. However, the fact that $\xi(T), \xi(T'') \in \repulbdr$ contradicts the strict decreasing property of $V(\xi(t))$ for $t \in \Xi$ (on the repulsive boundary $\repulbdr$, $\varphi$ attains the same value). The contradiction implies that the trajectory cannot go from the boundary $\repulbdr$ to the interior $\repulin$. Furthermore, the strict decreasing property of $V(\xi(t))$ also proves the repulsiveness of the boundary $\repulbdr$; more precisely, for every $\xi(0) \in \repulbdr$, we have $\xi(t) \notin \repulbdr$ for all $t>0$. Combining the previous arguments with the uniqueness of solutions shows that once the trajectory starts from the non-repulsive area $\repulex$, it will never reach the repulsive boundary $\repulbdr$ or the repulsive area $\repulin$. Thus the non-repulsive area $\repulex$ is indeed positively invariant.
\end{proof}

The almost repulsiveness of the closed repulsive area $\overline{\repulin}$ is stated in the following corollary.
\begin{coroll}[Almost repulsiveness of $\overline{\repulin}$] \label{cor2}
	The set $\overline{\repulin} \setminus \mathcal{W}(\rss)$ is repulsive. If the assumptions of Lemma \ref{lemma_inset} hold\footnote{In this case, replace $\set{P}$, $\phi$, $\pvf$, and $\pss$  with $\reactbdr$, $\varphi$, $\rvf$, and $\rss$ respectively in the lemma.}, then $\overline{\repulin}$ is almost repulsive.
\end{coroll}
\begin{proof}
	We need to show that if $\xi(0) \in \overline{\repulin} \setminus \mathcal{W}(\rss)$, then there exists $T>0$, such that $\xi(t) \notin \overline{\repulin}\setminus \mathcal{W}(\rss)$ for $t \ge T$. Since $\xi(0) \in \overline{\repulin} \setminus \mathcal{W}(\rss)$, we can use the same argument as in Lemma \ref{lem3} and conclude that there exists a time instant $T$, such that $\xi(T) \in \repulex$. Due to the positive invariance of $\repulex$, $\xi(t) \notin \overline{\repulin} \setminus \mathcal{W}(\rss)$ for $t \ge T$. If the assumptions of Lemma \ref{lemma_inset} hold, then the local inset $\set{W}(\rss)$ is of measure zero, so $\overline{\repulin}$ is almost repulsive by Definition \ref{def3}.
\end{proof}

The following lemma states that the absolute path-following error $|\phi(\xi(t))|$ is indeed bounded.
\begin{lemma}[Bounded path-following error] \label{lem4}
	If $\pss$ is bounded, then the absolute path-following error $|\phi(\xi(t))|$ of any trajectory $\xi(t)$ of \eqref{eqode} is bounded. 
\end{lemma}
\begin{proof}
	Suppose a trajectory of \eqref{eqode} is defined on the time interval $\mathcal{T}=[0, T_f]$, where $T_f \le \infty$. We want to prove that there exists a positive finite constant $M$ such that $|\phi(\xi(t)) \le M $ for all $t \in \mathcal{T}$. First, it is obvious that the absolute path-following error $|\phi(\cdot)|: \mbr[2] \to \mbr[]$ is continuous. Depending on where a trajectory $\xi(t)$ lies, three cases are discussed.
	
	\textbf{Case 1:} Suppose the trajectory $\xi(t)$ always lies in the closed reactive area $\overline{\reactin}$, then $|\phi(\xi(t))|$ attain its maximum value on the compact set $\overline{\reactin}$. Namely, $|\phi(\xi(t))| \le M_1$ for all $t \in \mathcal{T}$, where $ M_1 = \max_{p \in \overline{\reactin}} |\phi(p)|$.
	
	\textbf{Case 2:} Suppose the trajectory $\xi(t)$ always lies in the closed non-reactive area $\overline{\reactex}$, then the differential equation \eqref{eqode} is reduced to $\dot{\xi} = \npvf(\xi)$. By Lemma \ref{lemdic}, the trajectory either converges to the desired path $\set{P}$ or the singular set $\pss$. a) Suppose the trajectory $\xi(t)$ converges to the desired path $\set{P}$ as $t \to T_f$; that is, $|\phi(\xi(t))| \to 0$ as $t \to T_f$. Then, fixing $\epsilon>0$ such that $ |\phi(\xi(0))| > \epsilon$, there exists $0 < T' <T_f$, such that $|\phi(\xi(t))| <\epsilon$ for $t > T'$. Let $\gamma \defeq \max_{0 \le t \le T'} |\phi(\xi(t))|$, then the path-following error is bounded by $\max\{\epsilon, \gamma\}=\gamma=|\phi(\xi(0))|$, where the last equality is due to the decreasing property of the Lyapunov function $V=1/2 \, \phi^2(\xi)$, which will be elaborated in \textbf{Case 3} later. b) If the trajectory $\xi(t)$ converges to the set $\pss$ (it is naturally assumed that $\pss \subseteq \overline{\reactex}$, otherwise it is trivial), then there exists $0<T''<T_f$, such that $\dist(\xi(t), \pss)<\epsilon$ for $t > T''$. It follows that $\xi(t) \in \pss' \defeq \{ p \in \mbr[2] : \dist(p, \pss) \le \epsilon \}$ for $t > T''$. Since $\pss$ is bounded (hence compact), the set $\pss'$ is compact. Therefore, we can let $\beta \defeq \max_{x \in \pss'} |\phi(x)|$. Let $\alpha \defeq \max_{0 \le t \le T''} |\phi(\xi(t))|$, then the path-following error is bounded by $\max\{\alpha, \beta\}$. Overall, in both sub-cases a) and b), the path-following error $|\phi(\xi(t))|$ is bounded by $M_2 \defeq \max\{\gamma, \alpha, \beta\}$.
	
	\textbf{Case 3:} Suppose the trajectory lies alternately in $\overline{\reactex}$ and $\overline{\reactin}$. First, we suppose $\xi(0) \in \overline{\reactin}$. Denote all maximal connected closed intervals by $\rho_j \subseteq \mathcal{T}$, $j =1,2,\dots$, such that $\xi(t) \in \overline{\reactin}$ for $t \in \rho_j$. Similarly, denote all maximal connected closed intervals by $\delta_k \subseteq \mathcal{T}$, $k = 1,2,\dots$, such that $\xi(t) \in \overline{\reactex}$ for $t \in \delta_k$. For $t \in \bigcup_j \rho_j$, it follows that $|\phi(\xi(t))| \le M_1$ according to \textbf{Case 1}. Therefore, we only need to consider the time intervals $\delta_k$ over which the trajectory is in the closed non-reactive area $\overline{\reactex}$. For convenience, the minimum value in $\delta_k$ is denoted by $\delta_k^1$ (i.e., the first time instant). As \textbf{Case 2}, the differential equation \eqref{eqode} is reduced to $\dot{\xi} = \npvf(\xi)$. If $\mathcal{W}(\pss) \cap \reactbdr = \emptyset$, given $\xi(\delta_k^1) \in \reactbdr$, the trajectory will not converge to any points in $\pss$. Using the Lyapunov function candidate $V(\phi(\xi)) = 1/2 \; \phi^2(\xi)$ and taking the time derivative, one obtains $\dt[] V(\phi(\xi)) = -k_p \frac{1}{ \| \pvf(\xi) \| } \phi^2(\xi) \| \nabla \phi(\xi) \|^2$. This shows that $V(\phi(\xi(t)))$, hence $|\phi(\xi(t))|$ is decreasing as $t$ increases. Therefore, for each time interval $\delta_k$, $|\phi(\xi(t))|$ attains its maximum value $\max_{t \in \delta_k} |\phi(\xi(t))| = |\phi(\xi(\delta_k^1))| \le M_3$, where $M_3 \defeq \max_{p \in \reactbdr} |\phi(p)|$ (maximum attainable due to the compactness of the boundary $\reactbdr$). Therefore, for all $t \in \bigcup_k \delta_k$, $|\phi(\xi(t))| \le M_3 \le M_1$. Now we suppose $\xi(0) \in \overline{\reactex}$, then using the same analysis as before, it can be easily concluded that for all $t \in \bigcup_k \delta_k$, $|\phi(\xi(t))| \le \max\{M_1, d_0 \}$, where $d_0 = |\phi(\xi(0))|$. If the trajectory converges to the singular set $\pss$, then using the same analysis discussed in \textbf{Case 2}, the absolute path-following error $|\phi(\xi(t))|$ is bounded by $\max\{\alpha, \beta\}$. Overall, the path-following error $|\phi(\xi(t))|$ is bounded by $M_3 \defeq \max\{M_1, \alpha, \beta\}$. 
	To sum up, the path-following error $|\phi(\xi(t))|$ is bounded by $M \defeq \max\{M_1, \alpha, \beta, \gamma \}$.
\end{proof}
\begin{remark}
	Suppose a trajectory $\xi(t)$ does not converge to the singular set $\pss$, then the upper bound $M$ of the absolute path-following error $|\phi(\xi(t))|$ is only related to the initial error $\gamma=|\phi(\xi(0))|$ and the largest error between the reactive area $\overline{\reactin}$ and the desired path $\set{P}$ (i.e., $M_1= \max_{p \in \overline{\reactin}} |\phi(p)|$). Thus this upper bound can be reduced by starting a trajectory near the desired path or shrinking the reactive area. %
\end{remark}

In the sequel, the properties of the mixed area $\mathcal{M} = \repulex \cap \reactin$ will be investigated. First, one observes that the closed mixed area $\overline{\mathcal{M}}$ is not positively invariant. Since $\set{P} \not\subseteq\reactin$, there exists at least one vector on the reactive boundary which points from the boundary to the desired path. More precisely, there exists $p \in \reactbdr$ such that $\cvf(p) = \npvf(p)$ is not in the tangent cone\footnotemark of the closed mixed area $\overline{\mathcal{M}}$. By the Nagumo's theorem \cite[Theorem 4.7]{blanchini2008set}, the closed mixed area $\overline{\mathcal{M}}$ is not positively invariant. Thus there exists at least one point $\xi_0 \in \overline{\mathcal{M}}$ and $T>0$, such that the trajectory at time $T$ satisfies $\xi(T) \notin \overline{\mathcal{M}}$. However, the non-positive-invariance property of the closed mixed area is not sufficient to guarantee that any trajectory will not be trapped in this area. In fact, Lemma \ref{lem_limitation} demonstrates a common limitation of combining two vector fields. Recall the definition of the Poicar{\' e} index below. 

\begin{defn}[Poincar{\'e} index, {\cite[p. 68]{khalil2002nonlinear}}] \label{ChPre_index}
	Consider the second-order autonomous system $\dot{\xi}(t)=f(\xi(t))$, where $f: \mbr[2] \to \mbr[2]$ is continuously differentiable. Let $L$ be a \emph{simple closed} curve \emph{not} passing through any equilibrium point of the autonomous system. The orientation of the vector $f(x)$ rotates continuously as $x \in L$ traverses the curve $L$ in the counterclockwise direction, and upon returning to the original position, must have rotated an angle of $2k\pi$ for some integer $k$, where the angle is measured counterclockwise. The integer $k$ is called the \emph{index} of the closed curve $L$. If there is only a single isolated equilibrium point encircled by $L$, then $k$ is also called the \emph{index} of the equilibrium point.
\end{defn}
In a two-dimensional autonomous system, an equilibrium point of the system can be classified as a node, a focus, a center, or a saddle \cite[Chapter 2]{khalil2002nonlinear}. The next lemma determines the indices of these different types.
\begin{lemma}[{\cite[Lemma 2.3]{khalil2002nonlinear}}]~ 	\label{ChPre_index_thm}
	Consider $\dot{\xi}(t)=f(\xi(t))$, where $f: \mbr[2] \to \mbr[2]$ is continuously differentiable. Let $p \in \mathbb{R}^2$ be an equilibrium point of the system, and $L \subseteq \mbr[2]$ be a simple closed curve not passing through any equilibrium point.
	\begin{enumerate} [label=\arabic*), wide] 
		\item \label{item1} The index of $p$ is $+1$ if it is a node, a focus, or a center;
		\item \label{item2} The index of $p$ is $-1$ if it is a (hyperbolic) saddle;
		\item \label{item4} The index of $L$ is $0$ if it it does not encircle any equilibrium point;
		\item \label{item5} The index of $L$ is equal to the sum of the indices of the equilibrium points within it. 
		\item \label{item3} The index of a closed orbit is $+1$;
	\end{enumerate}
\end{lemma}

\begin{figure}[tb]
	\centering
	\includegraphics[width=0.5\linewidth]{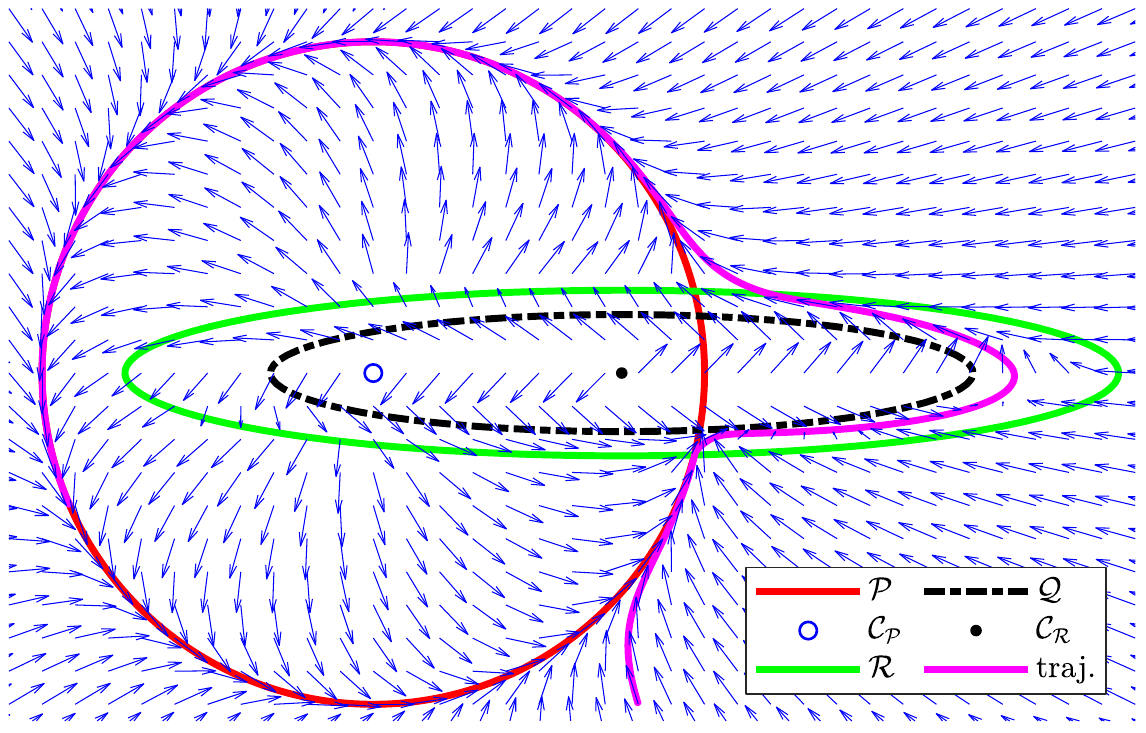}
	\caption{ In this example, the reactive area $\reactin$ is enlarged such that $\pss \cap \reactin \ne \emptyset$, and thereby Lemma \ref{lem_limitation} does not hold. Numerical calculation shows no equilibria in the mixed area $\mathcal{M}$, and thus $\css = \emptyset$ in \eqref{eq_css}.}
	\label{fig:no_saddle}
\end{figure}

\begin{lemma}[A common limitation] \label{lem_limitation}
	If $\pss \cap \reactin = \emptyset$, then there is at least one saddle point of \eqref{eqode} in the mixed area $\mathcal{M}$.
\end{lemma}
\begin{proof}
	On the repulsive boundary $\repulbdr$, by \eqref{eqrvf} and \eqref{eqcompvf_single}, the vector field $\cvf(\xi)$  is simplified to 	
	\begin{align*}
	\cvf(\xi) =\nrvf(\xi) &= \frac{1}{\| \rvf \|} \big( ( E - k_{r} \varphi(\xi) I) \nabla\varphi(\xi) \big) \\
	&\stackrel{\eqref{repulsive}}{=} \frac{1}{\| \rvf \|} \big(  E - k_{r} c I) \nabla\varphi(\xi) \big) \\
		&= \frac{\sqrt{k_r^2 c^2 + 1}}{\| \rvf \|} \cdot \underbrace{\frac{1}{\sqrt{k_r^2 c^2 + 1}} (E - k_r c I)}_{F} \cdot \nabla\varphi,
	\end{align*}
	where recall that $\varphi(\xi)=c$ for $\xi \in \repulbdr$ as shown in \eqref{repulsive}. Note that since $F F^\top = I$ and $\det(F) = 1$, $F$ is a rotation matrix. Therefore, for any point $p \in \mbr[2]$, $F \nabla \varphi(p)$ is a vector attained by rotating the gradient vector $\nabla \varphi(p)$ by some constant angle. In addition, the normalization of a vector field does not change the orientation of each vector. Thus, as a point $\xi$ traverses the boundary $\repulbdr$ in the counterclockwise direction, the vector $\nrvf(\xi)$ rotates the same angle as the gradient vector $\nabla \varphi(\xi)$ and the tangent vector $E \nabla \varphi(\xi)$ do. From the theorem of turning tangents \cite[p. 270]{do2016differential}, the tangent vector $E \nabla \varphi(\xi)$, hence the gradient vector $\nabla \varphi(\xi)$, rotates by an angle of $2\pi$ as a point $\xi$ traverses the boundary $\repulbdr$, which is a level curve of $\varphi$, in the counterclockwise direction. Therefore, from Definition \ref{ChPre_index}, the index of the boundary $\repulbdr$ is $1$; this also implies that there is at least one equilibrium point in $\repulin$ as shown by \ref{item5} of Lemma \ref{ChPre_index_thm}. Note that on the reactive boundary $\reactbdr$, the vector field is simplified to $\npvf(\xi)$, %
	and since $\pss \cap \reactin = \emptyset$, there are no equilibrium points (i.e., points in $\pss$) encircled by the reactive boundary $\reactbdr$. Then by \ref{item4} of Lemma \ref{ChPre_index_thm}, the index of the reactive boundary $\reactbdr$ is $0$. %
	As the composite vector field $\cvf$ is still continuous, we can conclude by Lemma \ref{ChPre_index_thm} that there must exist at least one saddle point in the mixed area $\mathcal{M}$. This is justified as follows. Denote the number of saddle points in the repulsive area $\repulin$ and the mixed area $\set{M}$ by $a_1 \ge 0$ and $a_2 \ge 0$, respectively, and denote the total number of nodes, foci and centers in the repulsive area $\repulin$ and the mixed area $\set{M}$ by $b_1 \ge 0$ and $b_2 \ge 0$, respectively. Since the index of the repulsive boundary $\repulbdr$ is $1$, we have $-a_1+b_1=1$ by \ref{item1}, \ref{item2}, \ref{item5} of Lemma \ref{ChPre_index_thm}. Similarly, since the index of the reactive boundary $\reactbdr$ is $0$, we have $-a_1+b_1-a_2+b_2=0$. The two equalities imply that $a_2=b_2+1 \ge 1$; namely, at least one saddle point exists in $\mathcal{M}$.
\end{proof}
\footnotetext{The tangent cone to a closed set $\mathcal{A} \subseteq \mbr[2]$ at a point $x \in \mbr[2]$ is defined as $\mathcal{T}_\mathcal{A}(x) = \{ v \in \mbr[2]: \liminf_{h \to 0} \dist(x + h v, \mathcal{A}) / h = 0 \}$.}
\begin{remark} \label{remark_limitationcounter}
	We impose the condition $\pss \cap \reactin = \emptyset$ here because this condition holds in many practical examples. For example, when the desired path is a straight line characterized by $\phi(x,y)=y$, then $\pss=\emptyset$ and the condition $\pss \cap \reactin = \emptyset$ automatically holds. This condition is also satisfied for all examples in Section \ref{sec_sim}. Moreover, the number of saddle points is exactly one more than the total number of nodes, foci and centers in the mixed area as shown in the proof. This is a topological limitation of the composite vector field regardless of the specific form of the reactive boundary, the repulsive boundary and the zero-in and zero-out functions. The best we can hope for is that there is only one saddle point in the mixed area, and thus trajectories starting from almost all initial conditions will \emph{not} be attracted to the saddle point. However, if $\pss \cap \reactin \ne \emptyset$, then it is possible that there are no equilibria in the mixed area $\mathcal{M}$ (see Fig. \ref{fig:no_saddle}), and thus this limitation can be removed. This is perhaps surprising since a similar topological limitation \emph{always} exists in traditional motion planning algorithms based on potential functions or navigation functions \cite{rimon1992exact,koditschek1990robot}. A similar topological obstruction is revealed in \cite{Braun2017OnE} in the context of merging control Lyapunov functions and control barrier functions. Specifically, \cite{Braun2017OnE} pinpoints the impossibility of globally asymptotically stabilizing the system origin using continuous controllers in environments with bounded obstacles. 
\end{remark}
	In the path-following problem (without obstacles), it is desirable that no singular points exist (i.e., $\pss = \emptyset$) such that global convergence to the desired path is guaranteed by Lemma \ref{lemdic}. In contrast, as implied by Lemma \ref{lem_limitation}, the emptiness of the singular set $\pss$ is not desirable when there are obstacles, since the condition $\pss \cap \reactin = \emptyset$ of Lemma \ref{lem_limitation} holds automatically, implying the existence of a saddle point in $\set{M}$.

We will prove that almost all initial conditions give rise to trajectories leaving the closed mixed area $\overline{\set{M}}$. The definition below explains what is meant by trajectories leaving a set.
\begin{defn}
	A trajectory $\xi: \mbr[] \to \mbr[2]$ \emph{leaves} a non-empty set $\mathcal{A} \subseteq \mbr[2]$ if there exists $t_1 > t_0$ such that $\xi(t_0) \in \mathcal{A}$ and $\xi(t_1) \notin \mathcal{A}$.
\end{defn}

The definition implies that whether a trajectory will enter the set $\mathcal{A}$ again (and remain or not remain there) is irrelevant. %
\begin{lemma}[Leaving $\overline{\mathcal{M}}$] \label{lem6}
	Suppose $\pss \cap \reactin = \emptyset$ and there is only one equilibrium point (i.e., a saddle point) $c_0 \in \css$ in the mixed area $\mathcal{M}$. Furthermore, suppose there exists a trajectory $\xi(t)$ starting from the repulsive boundary $\repulbdr$ and reaching the reactive boundary $\reactbdr$, then trajectories starting from almost all initial points in the closed mixed area $\overline{\mathcal{M}}$ will leave $\overline{\mathcal{M}}$.
\end{lemma}
\begin{proof}
	By Lemma \ref{lem_limitation}, the only equilibrium in the mixed area $\mathcal{M}$ is a saddle point. Thus the initial conditions of trajectories that converge to the saddle point form a set of measure zero. Since the index of the saddle point is $-1$, there are no closed orbits around it. Closed orbits could only be possible when they surround the repulsive area $\repulin$. However, since there is a trajectory moving from the repulsive boundary $\repulbdr$ to the reactive boundary $\reactbdr$, such closed orbits cannot exist (as there would be violation of the uniqueness of solutions). Therefore, trajectories starting from almost all initial points in the closed mixed area $\overline{\mathcal{M}}$ will leave $\overline{\mathcal{M}}$ by the Poicar\'e-Bendixson theorem \cite[Chapter 9]{wiggins2003introduction}.
\end{proof}

We are now ready to present the main theorem.
\begin{theorem} \label{thm1}
	The VF-CAPF problem with the vector field in \eqref{eqcompvf} is solved if the following conditions hold simultaneously:
	\begin{enumerate}[label=\textbf{C.\arabic*}, wide] 
		\item\label{cond1} $\mathcal{W}(\rss) \cap \repulbdr = \emptyset$, $\pss$ is bounded, and the initial condition $\xi(0) \notin \mathcal{W}(\pss)$;
		\item\label{cond2} $\pss \cap \reactin = \emptyset$ and there is only one equilibrium $c_0 \in \css$ in the mixed area $\mathcal{M}$;
		\item\label{cond3} There exists a trajectory $\xi(t)$ starting from the repulsive boundary $\repulbdr$ and reaching the reactive boundary $\reactbdr$.
	\end{enumerate}
\end{theorem}
\begin{proof}
	If there are no obstacles and given that $\xi(0) \notin \mathcal{W}(\pss)$, then due to Lemma \ref{lemdic}, the first control objective of the VF-CAPF problem is achieved. Given that $\mathcal{W}(\rss) \cap \repulbdr = \emptyset$, Lemma \ref{lem3} and Corollary \ref{cor2} imply that the second control objective of the VF-CAPF problem in Definition \ref{def1} is fulfilled. Since $\pss$ is bounded, Lemma \ref{lem4} shows that the path-following error is bounded, and the third control objective is met. Next, under conditions \ref{cond2} and \ref{cond3}, Lemma \ref{lem6} shows that almost all initial points in the closed mixed area $\overline{\mathcal{M}}$ give rise to trajectories leaving $\overline{\mathcal{M}}$. Due to the uniqueness of solutions, the vector field degenerates to the normalized path-following vector field $\cvf = \npvf$ once the trajectory leaves the reactive area. Then the trajectory will follow the desired path until it possibly returns to the reactive area again. Since $S(\xi)$ and $Z(\xi)$ are bounded, the norm of the composite vector field $\| \cvf \|$ is finite. Therefore, the time difference between two consecutive time instants $\Delta t_r>0$ of entry from the non-reactive area into the reactive area cannot be infinitely small. Thus the fourth control objective is accomplished.
\end{proof}
\begin{remark} \label{remark_original_vf}
	The technical results in this section still hold if one replaces the normalized vector fields $\npvf$ and $\nrvfi$ with the original ones $\pvf$ and $\rvfi$ in \eqref{eqcompvf}; i.e., \eqref{eqcompvf} is changed to 
	$
		\cvf(\xi) = \left( \prod_{i \in \mathcal{I}} \zinbump_{\repulbdr_i}(\xi) \right) \pvf(\xi) + \sum_{i \in \mathcal{I}} \left( \zoutbump_{\reactbdr_i}(\xi) \rvfi(\xi) \right).
	$
	This is because the core technical proofs rely on the Lyapunov analysis of the vector fields $\pvf$ or $\rvfi$ separately.  The proofs of the results for the new composite vector field with the original vector fields $\pvf$ and $\rvfi$ are almost the same except for minor changes; e.g., \eqref{eqdvdvarphi} needs to be multiplied by $\| \rvf \|$ without affecting the subsequent technical development. 
\end{remark}

The local insets $\mathcal{W}(\rss)$ and $\mathcal{W}(\pss)$  in Condition \ref{cond1} can be numerically calculated\footnote{For instance, if the singular points are hyperbolic, then the local insets are manifolds \cite[Theorem 7.6]{sastry2013nonlinear} that can be numerically computed via the graph transform method or the Lyapunov-Perron method \cite[Chapter 3.5]{wiggins2003introduction}}. Fortunately, the calculation can be avoided for some typical desired paths or boundaries, such as circles or ellipses, since the local insets are the same as the singular sets. \cmnt{As it is usually sufficient to model obstacles as ellipses, Condition \ref{cond1} can be greatly simplified.} More generally, sufficient conditions to avoid the calculation of $\set{W}(\cdot)$ are given in Corollary \ref{cor3}, which results from combining Lemma \ref{lemma_inset} and Theorem \ref{thm1}.

\begin{coroll} \label{cor3}
	Suppose the singular sets $\pss$ and $\rss$ are discrete, $\phi(p) H_{\phi}(p)$ and $\varphi(q) H_{\varphi}(q)$ have all negative eigenvalues for every point $p \in \pss$, $q \in \rss$. The VF-CAPF problem with the vector field in \eqref{eqcompvf} is solved if the conditions in Theorem \ref{thm1} hold, where $\set{W}(\rss)=\rss$ and $\set{W}(\pss)=\pss$.
\end{coroll}
\begin{remark} \label{remark7}
	Conditions \ref{cond2} and \ref{cond3} in Theorem \ref{thm1} might probably hold in practice, even though they are difficult to verify in theory. Condition \ref{cond2} might be satisfied by changing some design choices: a) the function $\phi$ characterizing the desired path $\set{P}$; b) the function $\varphi$ and the constant $c$ in \eqref{repulsive} characterizing the repulsive boundary $\repulbdr$ and reactive boundary $\reactbdr$, and c) the functions $S(\cdot)$ and $Z(\cdot)$ in the smooth zero-in and zero-out functions. \cmnt{In practice, a better design choice is to narrow down the mixed area $\mathcal{M}$ to reduce the uncertainties about trajectories in the mixed area.} Condition \ref{cond3} is not conservative because the existence of only one such trajectory is sufficient, but it is challenging to verify analytically. This condition is employed here to eliminate the possibilities of limit cycles in the mixed area, ensuring that trajectories can eventually leave the mixed area. Note that the proof of existence and non-existence of limit cycles in general is a challenging problem in nonlinear systems theory. There are only a few available tools, such as the Poincar\'e-Bendixson theorem, the Bendixson criterion and index theory \cite[Lemma 2.1-2.3]{khalil2002nonlinear}, \cite[Chapter 9]{wiggins2003introduction}. These tools might be used to verify condition \ref{cond3}. 
\end{remark}
The main disadvantages of the composite vector field approach discussed above are: a) Conditions \ref{cond2} and \ref{cond3} are difficult to check; b) In many cases, the limitation revealed in Lemma \ref{lem_limitation} exists. However, Conditions \ref{cond2} and \ref{cond3} are crucial to avoid the well-known phenomenon called \emph{deadlock}, but few algorithms in the literature provide a theoretical guarantee to avoid it \cite{zhou2017fast}. Nevertheless, we will use a \emph{switching vector field} in Section \ref{sec_switch} to replace these two conditions with easily-verifiable ones and also remove the limitation in Lemma \ref{lem_limitation}.

\begin{remark}[Moving obstacles] \label{remark_moving_obs}
In this remark, we will re-design the reactive vector field $\rvf': \mbr \times \mbr[2] \to \mbr[2]$ to include the ``motion information'' of obstacles, and thus obtain a new composite time-varying vector field $\cvf': \mbr \times \mbr[2] \to \mbr[2]$. Note that since the desired path is still static, the path-following vector field $\pvf:\mbr[2] \to \mbr[2]$ remains unchanged. Also note that since the new reactive vector field $\rvf'$ is time-varying, the normalization factor $1 / \| \rvf' \|$ of $\rvf'$ is also time-varying, and thereby the phase portrait of $\nrvf'$ is possibly different from that of the original vector field $\rvf'$ (c.f., Remark \ref{remark2}). For this reason, we do not use normalized vector fields in \eqref{eqcompvf}, but retreat to 
\begin{equation} \label{eqcompvf_original}
	\scalemath{0.85}{
		\cvf'(t, \xi) = \left( \prod_{i \in \mathcal{I}} \zinbump_{\repulbdr_i}(\xi) \right) \pvf(\xi) + \sum_{i \in \mathcal{I}} \left( \zoutbump_{\reactbdr_i}(t, \xi) \rvfi'(t, \xi) \right),
	}
\end{equation}
where we note that $\pvf$ and $\rvfi'$ are not normalized. Recall that the function $\varphi_i(t,\xi)$, which is related to the reactive boundary of the $i$-th obstacle, is twice continuously differentiable in $t$ (i.e., Assumption \ref{assump_smooth_varphi}). \cmnt{For simplicity, we only consider the \emph{translation} motion of obstacles, or equivalently, the translation motion of \emph{reactive boundaries}, of which \emph{the shapes are retained during the movement}.} Again by Assumption \ref{assump3}, it is sufficient to consider only one moving obstacle. \cmnt{If a robot's motion is guided by a vector field, it is natural to assume that its speed is always larger than that of the obstacle.} 

To ensure the previous results still hold for the new reactive vector field (hence the new composite vector field), we use a Lyapunov analysis to derive the new vector field $\rvf'$. Specifically, we add an additional time-varying term to the original reactive vector field \eqref{eqrvf} as follows:
\begin{equation} \label{eqnewvf}
	\rvf'(t, \xi) = E \nabla\varphi(t, \xi) - k_{r} \varphi(t, \xi) \nabla\varphi(t, \xi) + \varpi(t,\xi),
\end{equation}
where $\nabla\varphi(t, \xi) \defeq \frac{\partial \varphi(t,\xi)}{\partial \xi} \in \mbr[2]$ and $\varpi: \mathbb{R}_{+} \times \mbr[2] \to \mbr[2]$ is an additional term to be determined. We choose the Lyapunov function candidate $V(t, \xi) = \frac{1}{2} k \varphi^2(t,\xi)$, where $k$ is a positive constant. Now we have %
\begin{multline} \label{eqtmp1}
	\dt[]V(t,\xi) = \frac{{\rm d} V}{{\rm d} \varphi} \transpose{\nabla \varphi(t, \xi)} \dot{\xi} + \frac{{\rm d} V}{{\rm d} \varphi} \frac{\partial \varphi}{\partial t}   
	\stackrel{\eqref{eqnewvf}}{=} k \varphi \nabla \transpose{\varphi} \rvf' + \\ k \varphi \frac{\partial \varphi}{\partial t} 
	= -k k_r \varphi^2 \| \nabla \varphi \|^2 + k \varphi \left(\nabla \transpose{\varphi} \varpi + \frac{\partial \varphi}{\partial t} \right),
\end{multline}
where the second equation uses the equality $\dot{\xi} = \rvf'$ as we only consider the new reactive vector field $\rvf'$ now and conduct the Lyapunov analysis. Now let $\varpi$ be chosen as
\begin{equation} \label{eq14}
	\varpi = \frac{1}{\| \nabla \varphi \|^2} \left(-\frac{\partial \varphi}{\partial t} - l \varphi	\right) \nabla \varphi,
\end{equation}
where $l$ is a positive constant. Substituting \eqref{eq14} into \eqref{eqtmp1}, we have 
$
\dt[]V(t,\xi) = -k k_r \varphi^2 \| \nabla \varphi \|^2 -k l \varphi^2 \le -2 l V.
$
Therefore, by Theorem 4.9 in \cite{khalil2002nonlinear}, the integral curves of the new reactive vector field in \eqref{eqnewvf} will (globally) uniformly exponentially converge to the time-varying reactive boundary $\reactbdr$ characterized by $\varphi(t,\xi)=0$. 

We have shown above that the new reactive vector field $\rvf'$ can still guide trajectories to converge to the reactive boundary $\reactbdr$, which is time-varying now due to the motion of the obstacle. Therefore, by combining  \eqref{eqcompvf_original}, \eqref{eqnewvf} and \eqref{eq14}, we obtain a new composite vector field $\cvf'$. The previous results still hold for this new composite vector field $\cvf'$, for which the proofs are almost the same (see Remark \ref{remark_original_vf}). 
\end{remark}
\begin{remark}[Robustness] \label{remark_moving_robust}
	In practice, the time derivative information $\frac{\partial \varphi}{\partial t}$ in \eqref{eq14} is usually contaminated by measurement error. Fortunately, the exponential convergence property mentioned above provides some robustness against this error. We can still reduce the value of the Lyapunov function $V$ to an arbitrarily small positive value by increasing the gain $l$ in \eqref{eq14}. This is justified as follows. Let the gain $l > 1/2+\epsilon$, where $\epsilon>0$ is a constant. Suppose some time-varying measurement error $\rho(t,\xi) \in \mbr[]$ is added to the information of the time derivative $\frac{\partial \varphi}{\partial t}$. Therefore, \eqref{eq14} is perturbed as $\varpi = \frac{1}{\| \nabla \varphi \|^2} \left(-\frac{\partial \varphi}{\partial t} + \rho - l \varphi	\right) \nabla \varphi$, and \eqref{eqtmp1} becomes
	\begin{align}
		\dt[]V(t,\xi)  &= -k k_r \varphi^2 \| \nabla \varphi \|^2 - k l \varphi^2 + k \rho \varphi \label{eq22} \\
		&\le -\left( l-\frac{1}{2} \right)k \varphi^2 + \frac{1}{2} k \rho^2 \label{eq23} \\
		&\le - \epsilon k \varphi^2, \quad \forall |\varphi| \ge \alpha(|\rho|) > 0, \label{eq24}
	\end{align}
	where \eqref{eq23} has used the fact that $\rho \varphi \le \rho^2/2+\varphi^2/2$, and $\alpha(|\rho|)  = |\rho| / \sqrt{2l -2 \epsilon - 1} $ is a class $\kappa$ function. Therefore, by Theorem 4.9 in \cite{khalil2002nonlinear}, the system $\dot{\xi}(t)=\rvf'(t,\xi)$, where $\rvf'(t,\xi)$ is in \eqref{eqnewvf}, is input-to-state stable (ISS) w.r.t the measurement error $\rho$. If the measurement error is uniformly bounded $|\rho(\cdot)|<\rho_b$, where $\rho_0>0$ is a constant, this implies that by choosing $l > 1/2 + \epsilon$, the value of the Lyapunov function $V$ will eventually decrease to within $V=k \varphi^2/2 \le k \rho_b^2 / [ 2(2 l - 2\epsilon - 1) ] $. By choosing a large $\epsilon$, the decreasing rate of $V$ is greater as observed from \eqref{eq24}. Furthermore, one may assume that the measurement error is vanishing in the sense that $|\rho| \le \beta |\varphi|$, where $\beta>0$ is a constant. In this case, by choosing $l > \beta$, the Lyapunov function value $V$ will decrease (globally) uniformly exponentially to $0$. This is justified using the same argument as before. Namely, from \eqref{eq22}, we have $\dt[]V(t,\xi)  = -k k_r \varphi^2 \| \nabla \varphi \|^2 - k l \varphi^2 + k \rho \varphi \le -(l-\beta) k \varphi^2 = -2 (l-\beta)V$, and then we can employ Theorem 4.9 in \cite{khalil2002nonlinear}.%
\end{remark}

\section{Switching vector field} \label{sec_switch}
To replace the two conditions mentioned earlier and deal with undesirable equilibria in the mixed area, we introduce a switching vector field and prove that this switching vector field can solve the VF-CAPF problem. The idea of using a switching mechanism has also been adopted in \cite{braun2020explicit}, under the linearity assumption of the vector field.

Singular points in the mixed area $\set{M}$ appear where the components $S(\xi) \npvf(\xi)$  and $Z(\xi) \nrvf(\xi)$ in \eqref{eq_bump} cancel each other. Specifically, this happens only if $S(\xi)=Z(\xi)$. Namely, the extra singular points (should they exist) belong to the following set
\begin{equation} \label{eq_equalbump}
	\set{E} = \{ \xi \in \mbr[2] : S(\xi) = Z(\xi) \}.
\end{equation}
This set $\set{E}$ is non-empty since the functions $S(\cdot), Z(\cdot)$ are continuous, and $S(\cdot)$ decreases to $0$ ``radially inwardly'' towards the repulsive boundary $\repulbdr$, whereas $Z(\cdot)$ decreases to $0$ ``radially outwardly'' towards the reactive boundary $\reactbdr$. In addition, $\set{E}$ is compact and $\set{E} \subsetneq \set{M}$; i.e., this set does not intersect the reactive boundary $\reactbdr$ or the repulsive boundary $\repulbdr$. The functions $S(\cdot)$ and $Z(\cdot)$ are flexible design choices and characterize the set $\set{E}$. For simplicity of analysis, the functions $S(\cdot)$ and $Z(\cdot)$ are designed such that the set $\set{E}$ only contains ``rings''. More precisely, the set $\set{E}$ constitutes a finite number of one-dimensional compact connected submanifolds in $\mbr[2]$. In Section \ref{sec_sim}, we illustrate how to design these functions.%

The closed $\epsilon$-neighborhood $\set{E}^\epsilon$ of the set $\set{E}$ is
$
	\set{E}^\epsilon = \{\xi \in \set{M} : \dist(\xi, \set{E}) \le \epsilon \},
$
where $\epsilon>0$. Due to the compactness of $\set{E}$ and $\reactbdr$, and $\set{E} \cap \reactbdr = \emptyset$, one can always choose an $\epsilon>0$ sufficiently small such that\footnote{Since $\dist(\set{E},\reactbdr) \defeq \delta>0$, one can, for instance, choose $\epsilon = \delta/2$ such that $\dist(\set{E}^\epsilon, \reactbdr) >0$,  according to \cite[Chapter 1, Lemma 3.1]{stein2009real}.} 
$
	\dist(\set{E}^\epsilon, \reactbdr) >0.
$
The basic idea is that if a trajectory enters $\set{E}^\epsilon$, then it will possibly be attracted to a stable equilibrium in $\set{M}$. Nevertheless, we can then switch to another vector field such that this possibility is excluded. This new vector field is designed as a ``perturbed'' version of the reactive vector field in the sense that it is induced by a slightly enlarged reactive boundary, as introduced in the sequel. The existence of an enlarged reactive boundary is guaranteed by the following lemma. 
\begin{lemma}
		Given $\varphi$ and $\reactbdr$ in \eqref{reactive}, there exists a constant $\delta \ne 0$ such that $\reactbdr_\delta \defeq \varphi^{-1}(\delta)=\{\xi \in \mbr[2]: \varphi(\xi) = \delta \}$ is a one-dimensional compact connected submanifold in $\mbr[2]$, also satisfying $\dist(\reactbdr, \reactbdr_\delta)>0$ and $\reactin \subseteq \reactin_\delta$, where $\reactin_\delta$ is the \emph{perturbed reactive area} defined analogously to $\reactin$. 
\end{lemma}
\begin{proof}
	From Section \ref{sec_probf}, one knows that $0$ is a regular value of $\varphi$, and $\reactbdr=\varphi^{-1}(0)$ is a one-dimensional compact connected submanifold in $\mbr[2]$ (i.e., $\varphi^{-1}(0)$ is homeomorphic to $\mathbb{S}^1$). By the Ehresmann theorem \cite[p. 378]{cushman1997global}, which can be regarded as a generalization of the Morse theorem \cite[Theorem 2.6]{nicolaescu2011invitation}, there exists an open interval $\set{I}$ of $0$ such that $\varphi^{-1}(\set{I})$ is homeomorphic to $\set{I} \times \varphi^{-1}(0)$, given by a homeomorphism denoted by $\Gamma$, such that $\pi_{\set{I}} \circ \Gamma = \varphi |_{\varphi^{-1}(\set{I})} $, where $\pi_{\set{I}}$ is the projection onto the first factor. Therefore, for any $q \in \set{I}$, the level curve $\varphi^{-1}(q)$ is homeomorphic to $\reactbdr=\varphi^{-1}(0)$; namely, $\varphi^{-1}(q)$ is a one-dimensional compact connected submanifold in $\mbr[2]$ for any $q \in \set{I}$, and thus $\dist(\reactbdr, \reactbdr_q)>0$ for any $q \in \set{I}$, where $\reactbdr_q \defeq \varphi^{-1}(q)$. Moreover, choose an $\alpha \in \set{I}$ such that $\alpha \ne 0$ and $-\alpha \in \set{I}$. Let $\delta=\alpha$. If $\reactin \subseteq \reactin_\delta$ is satisfied, then the proof is complete; otherwise, we choose $\delta=-\alpha$. 
\end{proof}
	This lemma means that the shape of the level set $\varphi^{-1}(\delta)$ is similar to the zero-level set $\varphi^{-1}(0)=\reactbdr$. \cmnt{More precisely, $\varphi^{-1}(\delta)$ is homeomorphic to the reactive boundary $\reactbdr$.} For example, suppose the reactive boundary enclosing an obstacle is an ellipse, characterized by $\varphi=x^2/a^2+y^2/b^2-1=0$, where $a,b \ne 0$ are constants. \emph{All} the level sets $\varphi^{-1}(k)$ for $k>-1$ are ellipses of different sizes.

\begin{figure}[tb]
	\centering
	\begin{tikzpicture}[scale=1]
	\tikzstyle{mydash}=[thick,dashed]
		
	\node[xshift=0,yshift=-0.1] at (0,0) {	\includegraphics[width=\columnwidth]{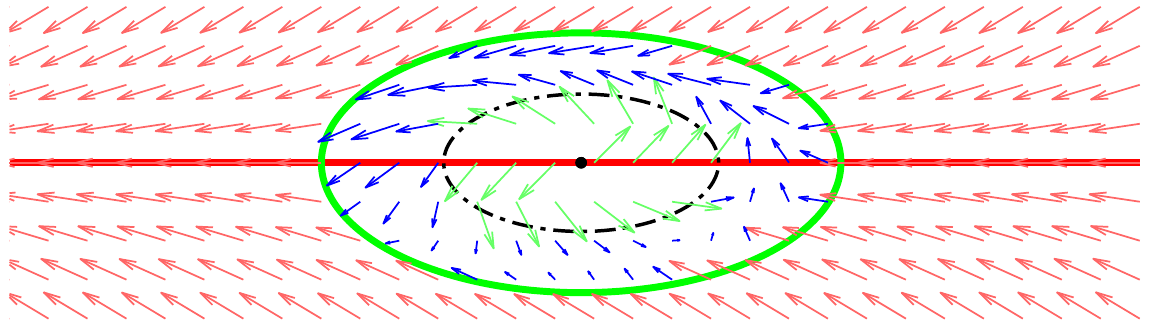}};

	\draw[red,thick] (-4.2,0) node[below, outer sep=2pt, xshift=.5cm] {$\set{P}$};	%
	\draw[gray!30, fill=gray!30,opacity=0.8] (-2.2, 0) circle[radius=22pt] node[above,xshift=-0.3cm,black] {$\set{J}_{\rm int}^{\rm o,\epsilon_{o}}$}; %
	\draw[red, fill=red] (-2.2, 0) circle[radius=1.5pt]; %
	\draw[-latex, thick, blue, mydash] (-2.2, 0) -- (-3.1,0); %
	
	\draw[green] (1.6,-0.35) node {$\reactbdr$} ; %
	\draw[color=green, style=mydash] (0,0) circle[x radius=2cm, y radius=1cm, scale=1.1] node at(2.1,-0.8) {$\reactbdr_\delta$};	%
	\draw[black] (-0.6,0.2) node {$\repulbdr$} ; %

	\draw[line width=1.8mm,color=gray,opacity=0.6] (0,0) circle[x radius=2cm, y radius=1cm, scale=0.7]  node at(1.7,0.2) {$\set{E}^\epsilon$};	%
	\draw[color=black] (0,0) circle[x radius=2cm, y radius=1cm, scale=0.7];		%
	\fill[fill=red!50, opacity=1,scale=0.13,opacity=0.8] plot[smooth cycle] coordinates {(-4,-0.5) (-2.5,-2.5) (1,-2) (4,-2.4) (4.6, 0) (4, 1.6) (0,3) (-2.5, 1.5) (-3, 1)};		 %

	\draw[-latex, thick]  (0.7,-1.3) ..controls (0.4, -1) .. (-0.5,-0.75);
	\draw[-latex, thick, gray]  (-0.5,-0.75) parabola  (-1.7,-0.6);
	\draw[latex-, thick]  (-2.8,-0.1) parabola  (-1.7,-0.6);
	\draw[fill=black] (-0.58, -0.64) circle[radius=1.5pt];
	\draw[fill=white] (0.9, -0.54) circle[radius=1.5pt];
	\draw[fill=white] (-1.35, -0.2) circle[radius=1.5pt];
	\end{tikzpicture}
	\caption{Concepts for the switching vector field \eqref{eqswitch}. The pink irregular shape is the obstacle. The dot-dashed black line is the repulsive boundary $\repulbdr$. The solid black line is $\set{E}$ and the shaded area around $\set{E}$ is $\epsilon$-neighborhood $\set{E}^\epsilon$. Two white points on the solid line $\set{E}$ are saddle points, and the black point is a stable equilibrium of \eqref{eqode}.  The solid green line and dashed green line are the reactive boundary $\reactbdr$ and the perturbed reactive boundary $\reactbdr_\delta$, respectively. The red horizontal solid line is the desired path $\set{P}$. The red point is an intersection point between $\reactbdr_\delta$ and $\set{P}$, where a blue arrow represents the outward-pointing normal of $\reactbdr_\delta$. The gray disk symbolizes the set $\set{J}_{\rm int}^{\rm o,\epsilon_{o}}$ in \eqref{eq_jintepsilon}. Three arrows compose a trajectory starting from beyond the reactive area, where the black and gray arrows correspond to $\sigma=1$ and $\sigma=2$ respectively. }
	\label{fig:switching}
\end{figure}
Note that the only difference in the definition of $\reactbdr_\delta$ from that in \eqref{reactive} is the constant $\delta$. Therefore, we can similarly derive a \emph{perturbed reactive vector field} $\rvfp$ by replacing the surface function $\varphi$ with $\varphi^{\delta} \defeq \varphi-\delta$ in \eqref{eqrvf}. This notation also implies that $\varphi^{0}=\varphi$. Accordingly, we define the singular set $\rssp$ corresponding to this perturbed reactive vector field $\rvfp$ as $\rssp \defeq \{\xi \in \mbr[2] : \rvfp(\xi) =0 \} = \{\xi \in \mbr[2] : \nabla \varphi^{\delta}(\xi) =0 \}$. Since $\varphi^{\delta} = \varphi-\delta$, we have $\nabla \varphi^{\delta} = \nabla \varphi$ and $\rssp=\rss$.

We define $\set{J}_{\rm int} \defeq \reactbdr_\delta \cap \set{P} \ne \emptyset$, which is the set of all intersection points between the perturbed reactive boundary $\reactbdr_\delta$ and the desired path $\set{P}$. For simplicity, we assume that this set $\set{J}_{\rm int}$ is finite, as otherwise, one can choose a different perturbed reactive boundary $\reactbdr_\delta$ such that this assumption holds. The \emph{outward-pointing normal} \cite{tapp2016differential} of the perturbed reactive boundary $\reactbdr_\delta$ at a point $q \in \reactbdr_\delta$, denoted by $N_{\rm o}(q) \in \mathbb{S}^1 \subseteq \mbr[2]$, is an outward-pointing unit vector perpendicular to $\reactbdr_\delta$ at $q \in \reactbdr_\delta$. Now we can define the set of points where the path-following vector field $\pvf$ points towards a similar direction to the outward-pointing normal of $\reactbdr_\delta$ at the intersection points:
$
	\set{J}_{\rm int}^{\rm o} \defeq \{\xi \in \set{J}_{\rm int} : \npvf(\xi)^\top N_{\rm o}(\xi) > 0  \},
$
which is a finite set. Starting from any point of this set, the integral curves of the path-following vector field will be ``driven out'' of the perturbed reactive area $\reactin_\delta$. Due to the continuity of $\npvf$ and the finiteness of $\set{J}_{\rm int}^{\rm o}$, one can choose an $\epsilon_{o}>0$ sufficiently small such that for every two distinct intersection points $q_i, q_j \in \set{J}_{\rm int}^{\rm o}$, their closed $\epsilon_{o}$-neighborhoods are disjoint (i.e., $q_i^{\epsilon_{o}}\cap q_j^{\epsilon_{o}} = \emptyset  $, where $q_i^{\epsilon_{o}} \defeq \{\xi \in \mbr[2]: \dist(\xi, q_i) \le \epsilon_{o} \}$ and $q_j^{\epsilon_{o}}$ is defined similarly), and in each of these $\epsilon_{o}$-neighborhood, $\npvf$ points towards a similar direction to the outward-pointing normal at the corresponding intersection point; namely, $\forall q \in \set{J}_{\rm int}^{\rm o}$ and $\forall \xi \in q^{\epsilon_{o}}$, there holds $\npvf(\xi)^\top N_o(q) > 0$. We define the union of these $\epsilon_{o}$-neighborhoods to be
\begin{equation} \label{eq_jintepsilon}
	\set{J}_{\rm int}^{\rm o,\epsilon_{o}} \defeq \bigcup_{q \in \set{J}_{\rm int}^{\rm o}} q^{\epsilon_{o}}.
\end{equation}
See Fig. \ref{fig:switching} for the introduced new concepts.

\begin{figure}[tb]
	\centering
	\begin{tikzpicture}[scale=0.65, every node/.style={scale=0.65}]
	\tikzset{
		->, %
		>=stealth, %
		node distance=4cm, %
		every state/.style={thick, fill=gray!10}, %
		initial text=$ $, %
	}
	\node[state] (q1) {$\begin{array}{c} \sigma=1 \\ (\chiup_1 = \cvf) \end{array}$};
	\node[state, right of=q1, scale=0.9] (q2) {$\begin{array}{c} \sigma=2 \\ (\chiup_2 = \rvfp) \end{array}$};
	\draw (q1) edge[bend left, above] node{$\xi \in \set{E}^\epsilon$} (q2)
	(q2) edge[bend left, below] node{$\xi \in \reactex \cap \set{J}_{\rm int}^{\rm o,\epsilon_{o}}$} (q1);
	\end{tikzpicture}
	\caption{The discrete transitions of the switching signal $\sigma$ in \eqref{eqswitch}.}
	\label{fig:sigma}
\end{figure}
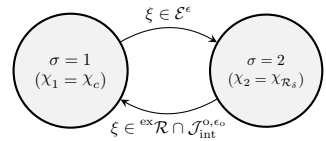

Now we consider the following switching system
\begin{equation} \label{eqswitch}
	\dot{\xi}(t) = \chiup_{\sigma(t)}(\xi(t)),
\end{equation}
where $\sigma: [0, \infty) \to \{1,2\}$ is the \emph{switching signal} \cite[p. 6]{liberzon2003switching}, of which the discrete transitions depend on its previous discrete state $\lim_{\tau \to t^-} \sigma(\tau)$ and the continuous state $\xi(t)$ at time $t$, as shown in Fig. \ref{fig:sigma}. More precisely, $\sigma(t)=1$ if $\lim_{\tau \to t^-} \sigma(\tau)=2$ and $\xi(t) \in  \reactex \cap \set{J}_{\rm int}^{\rm o,\epsilon_{o}}$, and the right-hand side of \eqref{eqswitch} becomes the composite vector field $\chiup_1 = \cvf$; $\sigma(t)=2$ if $\lim_{\tau \to t^-} \sigma(\tau)=1$ and $\xi(t) \in \set{E}^\epsilon$, and the right-hand side of \eqref{eqswitch} becomes the perturbed reactive vector field $\chiup_2 = \rvfp$; otherwise, the switching signal retains the previous value: $\sigma(t) = \lim_{\tau \to t^-} \sigma(\tau)$. Solutions to \eqref{eqswitch} are interpreted \emph{in the sense of Carath{\'e}odory} \cite[p. 10]{liberzon2003switching}.\cmnt{; that is, a solution to \eqref{eqswitch} is an absolutely continuous function $\xi: [0, \infty) \to \mbr[n]$  satisfying $\xi(t) = \xi(t_0) + \int_{t_0}^{t} \chiup(\tau, \xi(\tau)) d\tau $, where $\chiup(\tau, \xi(\tau))$ in the integral is adapted from the right-hand side of \eqref{eqswitch} to show its explicit dependence on time $t$ due to the switching signal $\sigma(t)$. The solution is piecewise differentiable and satisfies the differential equation \eqref{eqswitch} almost everywhere. }

Recall that Conditions \ref{cond2} and \ref{cond3} in Theorem \ref{thm1} are used to prevent trajectories from converging to an attractive equilibrium in $\css \subseteq \set{M}$ and from converging to a closed orbit in $\set{M}$ respectively. Namely, the common objective is to prevent trajectories from getting stuck in the mixed area $\set{M}$. Using the switching vector field $\chiup_{\sigma(t)}$, we can replace these two conditions with more verifiable ones in the following theorem.

\begin{theorem} \label{thm2}
	Suppose the functions $S(\cdot)$ and $Z(\cdot)$ in \eqref{eq_bump} are chosen such that the set $\set{E}$ in \eqref{eq_equalbump} is a one-dimensional connected manifold\footnote{Roughly speaking, $\set{E}$ is (the trace) of a simple closed curve.}. Consider the switching system \eqref{eqswitch}, the VF-CAPF problem is solved if the following conditions hold simultaneously:
	\begin{enumerate}[label=\textbf{C.\arabic*}, wide] 
		\item\label{cond1_sw} Condition \ref{cond1} in Theorem \ref{thm1}; i.e., $\mathcal{W}(\rss) \cap \repulbdr = \emptyset$, $\pss$ is bounded, and the initial condition $\xi(0) \notin \mathcal{W}(\pss)$;
		\item\label{cond2_sw} $\set{W}(\rssp) \cap \set{E}^\epsilon = \emptyset$; 
		\item\label{cond3_sw} The initial conditions $\xi(0) \notin \reactin$ and $\sigma(0)=1$.
	\end{enumerate}
	In particular, the Zeno behavior does not occur.
\end{theorem}
\begin{proof}
	\emph{Step 1:} We first show that trajectories of \eqref{eqswitch} will not converge to any equilibrium points in $\set{E}$. Precisely, trajectories of \eqref{eqswitch} will not converge to the set $\set{S} \defeq \css \cap \set{E} = \css \cap \set{E}^\epsilon$, where $\css$ defined in \eqref{eq_css} is the set of singular points of the composite vector field $\cvf$, which happens to be the set of equilibrium points of \eqref{eqswitch} when $\sigma=1$. Note that the perturbed reactive vector field $\rvfp$ is activated only if $\xi(t) \in \set{E}^\epsilon$, but given condition \ref{cond2_sw} and the dichotomy convergence property (Lemma \ref{lemdic}), trajectories of $\dot{\xi}=\chiup_{2}(\xi) = \rvfp(\xi)$ will not converge to $\set{S}$. Therefore, it suffices to consider only the case $\sigma=1$. Suppose there exists a trajectory $\xi(t)$ of \eqref{eqswitch}, where $\chiup_{\sigma=1}=\cvf$, converging to $\set{S}$; then there exists a time instant $T_1 \ge 0$ such that $\xi(T_1) \in \set{E}^\epsilon$. In this case, the system switches to $\dot{\xi}=\rvfp(\xi)$. However, as mentioned before, the trajectory will then converge to the perturbed reactive boundary $\reactbdr_\delta$ if no switching happened afterward. In particular, there exists $T_2 > T_1$ such that $\xi(T_2) \in \reactbdr$. Since $\dist(\set{E}, \reactbdr) > 0$, this implies that the trajectory cannot converge to $\set{S}$.%
	
	\emph{Step 2:} Now we show that there are no closed orbits in the reactive area $\reactin$ (given condition \ref{cond3_sw}). Using similar arguments as before, it suffices to investigate only the case when $\sigma=1$ and the switching system \eqref{eqswitch} becomes $\dot{\xi} = \cvf(\xi)$. As shown in the proof of Lemma \ref{lem_limitation}, the index of the reactive boundary $\reactbdr$ is $0$. Since there are are no equilibrium points between the reactive boundary $\reactbdr$ and $\set{E}$ (precisely, $\reactin \cap \prescript{{\rm ex}}{}{\set{E}}$, where $\prescript{{\rm ex}}{}{\set{E}}$ is defined analogously to $\reactex$), any closed orbit starting between the reactive boundary $\reactbdr$ and $\set{E}$ must intersect $\set{E}^\epsilon$ to enclose equilibria such that its index becomes $1$ \cite[Lemma 2.3]{khalil2002nonlinear}. Therefore, any trajectory corresponding to a closed orbit  must intersect $\set{E}^\epsilon$, but then the vector field switches to $\rvfp$. Therefore, there cannot be closed orbits in the reactive area $\reactin$.
	
	\emph{Step 3:} Now we show that trajectories will eventually leave the closed reactive area $\overline{\reactin}$. \emph{Step 1} has proved that any trajectories cannot converge to the set $\set{E}$, since any such trajectories would be driven to move to the reactive boundary $\reactbdr$ by the perturbed reactive vector field $\chiup_{\sigma=2}=\rvfp$. If the switching signal does not change to $\sigma=1$, then the trajectory will converge to the perturbed reactive boundary $\reactbdr_\delta$ according to condition \ref{cond2} and the dichotomy convergence property in Lemma \ref{lemdic}. Since the switching signal only switches to $\sigma=1$ if $\xi(t) \in \reactex \cap  \set{J}_{\rm int}^{\rm o,\epsilon_{o}}$, this implies that the trajectory will eventually leave the closed reactive area $\overline{\reactin}$. Moreover, once the condition $\xi(T) \in \reactex \cap  \set{J}_{\rm int}^{\rm o,\epsilon_{o}}$ is satisfied at some time $T$, the vector field becomes $\chiup_{\sigma=1}=\cvf=\npvf$. Due to the property of the set $\set{J}_{\rm int}^{\rm o,\epsilon_{o}}$ in \eqref{eq_jintepsilon}, the trajectory will eventually leave the perturbed reactive boundary $\reactbdr_\delta$.

	\emph{Step 4:} Finally, we show that the Zeno behavior cannot occur. Let $d \defeq \dist(\set{E}^\epsilon, \reactbdr)>0$, and $v_m \defeq \max_{\xi \in \overline{\set{M}}} \{ \chiup_{\sigma=1}, \chiup_{\sigma=2} \} < \infty$. Therefore, the duration $\Delta t$ between any two switching time instants is lower bounded by $d/ v_m > 0$, thus excluding the Zeno behavior.
\end{proof}

\begin{remark}
	One can observe that in this switching approach, the composite vector field $\cvf$ degenerates to the normalized path-following vector field $\npvf$, and the switching happens between $\npvf$ and $\rvfp$. Thus the design of the composite vector field seems redundant. However, the switching mechanism is only introduced to solve the deadlock problem. Once there is no such problem, then a composite vector field is more efficient as it can avoid the non-smooth and possible tortuous motion caused by the switching of different vector fields.
\end{remark}

Similarly to Corollary \ref{cor3}, the following corollary facilitates the verification of the conditions in Theorem \ref{thm2}. Note that since $\varphi^\delta=\varphi-\delta$, the Hessian matrices satisfy $H_{\varphi}(\cdot)=H_{\varphi^{\delta}}(\cdot)$. 
\begin{coroll} \label{cor4}
	Consider the switching system \eqref{eqswitch} and suppose the singular sets $\pss$ and $\rss=\rssp$ are discrete. If $\phi(p) H_{\phi}(p)$, $\varphi(q) H_{\varphi}(q)$ and $\varphi^{\delta}(q) H_{\varphi}(q)$ have all negative eigenvalues\footnote{If $|\delta|$ is sufficiently small (i.e., $|\delta| \approx 0$), then the signs of eigenvalues of $\varphi(q) H_{\varphi}(q)$ and $\varphi^{\delta}(q) H_{\varphi}(q)$ for every point $q \in \rss$ are the same. Therefore, one does not need to check the condition related to $\varphi^{\delta}(q) H_{\varphi}(q)$.} for every point $p \in \pss$ and $q \in \rss$, where $\varphi^{\delta} \defeq \varphi-\delta$, then the VF-CAPF problem is solved if conditions in Theorem \ref{thm2} hold, where $\set{W}(\rss)=\set{W}(\rssp)=\rss$ and $\set{W}(\pss)=\pss$.
\end{coroll}
\begin{remark}
	We have shown that the switching vector field discussed in this section can overcome the common limitation stated in Lemma \ref{lem_limitation}. In fact, many existing studies in the literature have adopted hybrid controllers. The paper \cite{Marley2021SynergisticCB} proposes a synergistic control barrier function that enables switching between overlapping non-hybrid control barrier functions to guarantee collision avoidance. The algorithm is still effective for nonholonomic vehicles. Another hybrid state feedback control law is studied in \cite{Sanfelice2006RobustHC}, which achieves robust global regulation to a target while avoiding an obstacle. The work also provides a fundamental result showing the impossibility of using pure state feedback to accomplish the same task. These two works have not yet taken into account complicated robot motion tasks, such as path following. In \cite{Casau2019AdaptiveBO}, a hybrid controller induced by a synergistic Lyapunov function and feedback pair is designed for obstacle avoidance. For multi-vehicle systems, \cite{Poveda2021RobustCH} presents a class of model-free hybrid controllers to achieve robust source seeking and obstacle avoidance simultaneously. These two studies only consider one obstacle with a known location. 
\end{remark}

\begin{figure}[tb]
	\centering
	\includegraphics[width=0.6\linewidth]{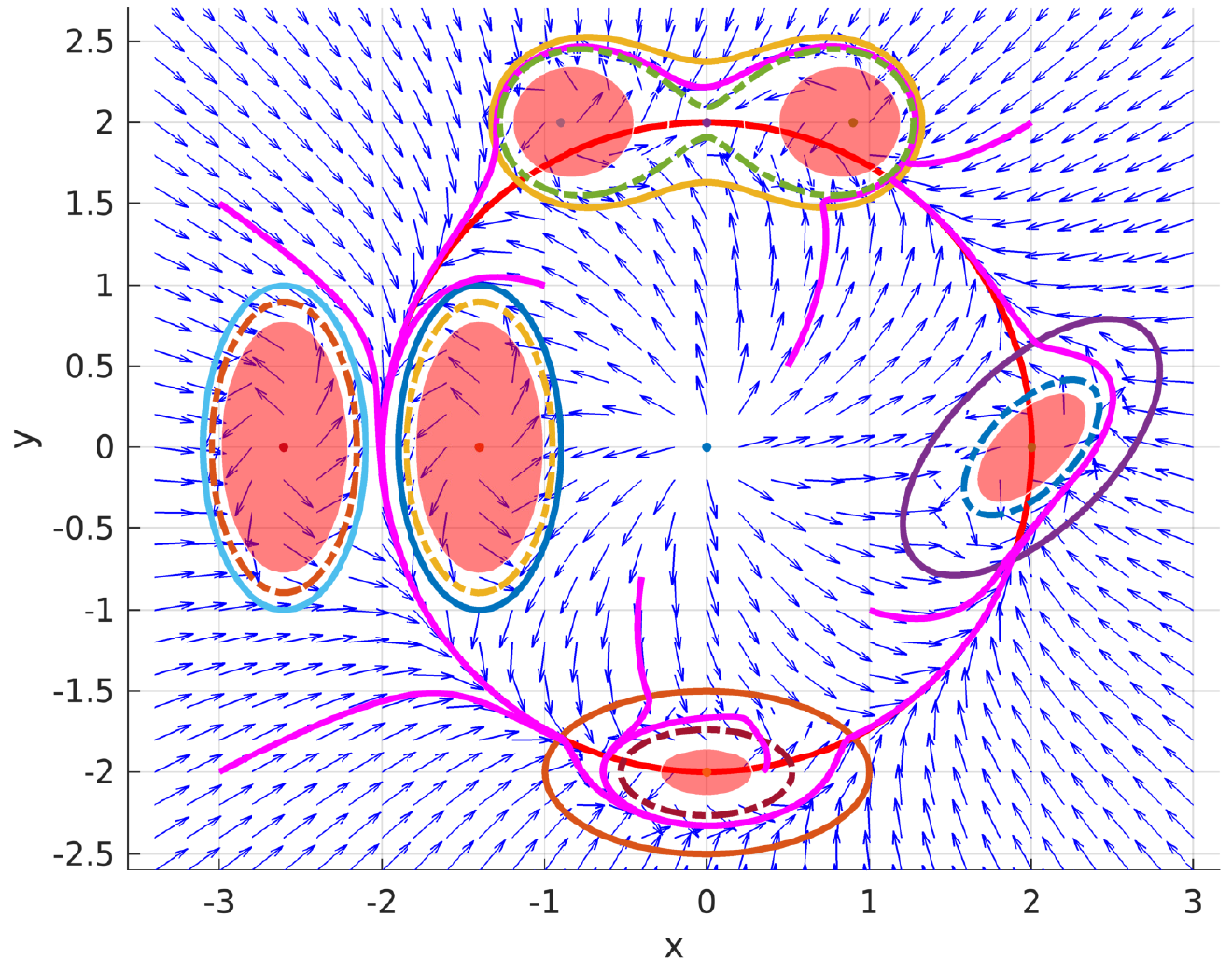}
	\caption{Simulation I. The red curve (partly covered by the pink curve) is the desired path. The red transparent shapes are obstacles, the solid closed curves enclosing them are the reactive boundaries and the dashed lines are the repulsive boundaries. The pink curves are trajectories starting from eight different positions. One of the initial point is in the repulsive area. }
	\label{fig:sim2}
\end{figure}
\section{Simulations} \label{sec_sim}
\subsection{Smooth zero-in and zero-out functions}
Given the function $\varphi$ and a constant $c$ as in \eqref{repulsive}, we define two \emph{smooth} functions:
\begin{equation*} 
\scalemath{0.78}{
f_1(\xi) = 
\begin{dcases}
0 & \varphi(\xi) \le c \\
\exp \left( \frac{l_1}{c - \varphi(\xi)} \right) & \varphi(\xi) > c,
\end{dcases}
\quad
f_2(\xi) = 
\begin{dcases}
\exp \left( \frac{l_2}{ \varphi(\xi) } \right) & \varphi(\xi) < 0  \\
0 &\varphi(\xi) \ge 0 ,
\end{dcases}
}
\end{equation*}
where $l_1>0$, $l_2>0$ are used to change the decaying or increasing rate. Throughout this section, the zero-in and zero-out bump functions are designed as %
\begin{equation} \label{eq_zeroinoutbump}
\zinbump_{\repulbdr}(\xi) =  \frac{f_1(\xi)}{f_1(\xi)  + f_2(\xi)}, \quad 
\zoutbump_{\reactbdr}(\xi) =  \frac{f_2(\xi)}{f_1(\xi) + f_2(\xi)}.
\end{equation}   
The denominators of the above bump functions are always positive, and the functions satisfy the conditions in Corollary \ref{cor1}. In addition, thanks to this design, the set $\set{E}$ in \eqref{eq_equalbump} is simply the $l_2 c / (l_1 + l_2)$-level set of $\varphi$; i.e., $\set{E} = \varphi^{-1} ( l_2 c / (l_1 + l_2)  ).$
\cmnt{If appropriate functions $\phi$ and $\varphi$ are chosen, then $\set{E}$ has the same geometric shape as $\reactbdr=\varphi^{-1}(0)$. }

\subsection{Static obstacles and single-integrator model in 2D}
In this simulation example, the desired path is a circle described by the function canonically chosen as $\phi(x,y) = x^2 + y^2 -R^2=0$, where $R > 0$ is the radius. According to Lemma \ref{lemma_inset}, one can conclude that $\pss=\mathcal{W}(\pss)=\{0\}$. Six obstacles are placed either directly on the desired path, or very close to it (see Fig. \ref{fig:sim2}). Depending on the shapes of the obstacles, some reactive boundaries are described by rotated ellipses: $\varphi(x,y) =   ((x-o_x) \cos\beta + (y-o_y)\sin\beta )^2 / a^2 + ((x-o_x)\sin\beta - (y-o_y)\cos\beta )^2 / b^2 - 1 = 0$, 
where $a,b>0$ and $\beta$ is the rotation angle about the center of the ellipse $(o_x, o_y)$. The singular sets and the corresponding insets are simply $\rss=\mathcal{W}(\rss)=\{(o_x, o_y)\}$. Two obstacles are enclosed by one reactive boundary which is modeled by a Cassini oval described by $\varphi(x,y)=[(x-0.9)^2+(y-2)^2][(x+0.9)^2+(y-2)^2] - 0.9 = 0$, and we have $\rss=\mathcal{W}(\rss)=\{(\pm0.9, 2)\}$ for this reactive boundary. Since the functions $\phi$ and $\varphi$ chosen for circles, ellipses and Cassini ovals are common in practice, we call them \emph{canonical functions} for simplicity.

As shown in Fig. \ref{fig:sim2}, starting from eight different positions, $(1,-1)$, $(2,2)$, $(-1, 1)$, $(-3, 1.5)$, $(-3, -2)$, $(0.35, -2)$, $(0.5,0.5)$ and $(-0.4,-0.8)$, all trajectories successfully follow the desired path and bypass the obstacles without entering the repulsive areas (except when starting from the repulsive area). Note that the trajectories smoothly pass the narrow passage surrounded by two vertical ellipses, while the vanilla artificial potential field method can hardly achieve this \cite{koren1991potential}. Also, note that this method is effective even though the Cassini oval is \emph{not} convex. By numerical calculations, we find that there is only one saddle point in $\set{E}$ in each reactive area. Therefore, as verified by the simulation, no trajectories are attracted to stable points in $\set{E}$, so the switching vector field is not employed.

\subsection{Switching vector field to overcome the deadlock}
\begin{figure}[tb]
	\centering
	\begin{tikzpicture}
	\node[anchor=south west,inner sep=0,xshift=-1.05cm,yshift=-0.75cm,scale=0.7] at (0,0) {\includegraphics[width=\linewidth]{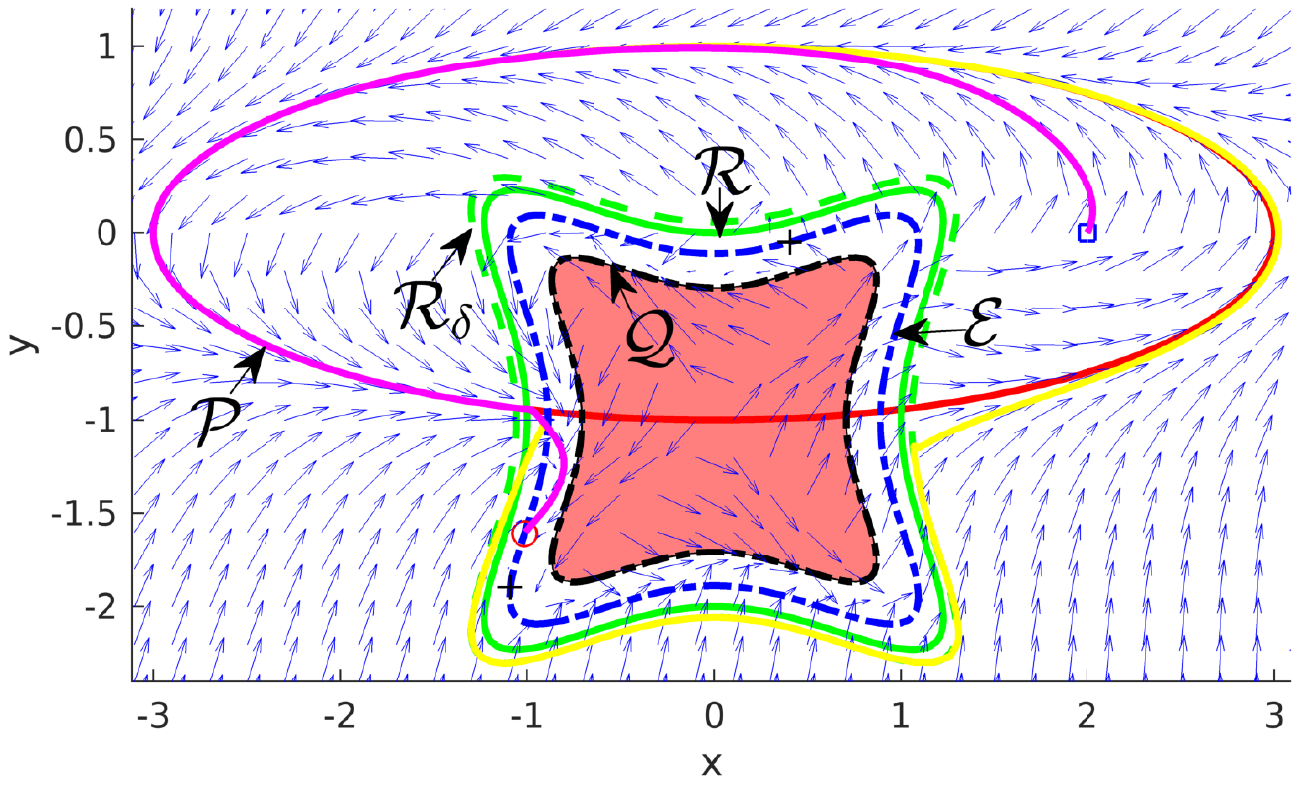}};

	\tikzset{
		cross/.pic = {
			\draw[rotate = 45] (-#1,0) -- (#1,0);
			\draw[rotate = 45] (0,-#1) -- (0, #1);
		}
	}
	
	\draw[color=red, ultra thick] (1.46, 0.46) circle[radius=1.5pt] ; %
	\path (1.4, 0.21) pic[red, ultra thick,rotate=45] {cross=2.5pt};	%
	\path (2.7, 1.85) pic[red, ultra thick,rotate=45] {cross=2.5pt};	%
	\end{tikzpicture}
	\caption{Simulation II. The red transparent area represents an obstacle. Two red crosses represent two saddle points and the red small circle represent the stable equilibrium in $\set{E}$. The magenta curve is a trajectory of \eqref{eqcompvf} starting from $(2,0)$. It  is attracted to the stable equilibrium and gets stuck in the reactive area. This deadlock behavior does not occur when the trajectory (yellow curve) is generated by the switching vector field.  }
	\label{fig:swtichvf}
\end{figure}
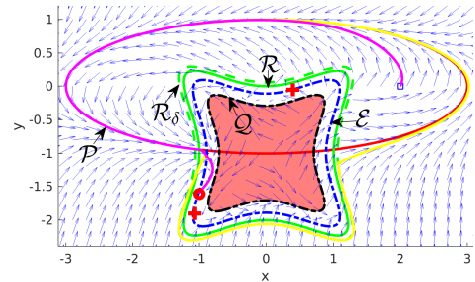

In this simulation, we use one obstacle for simplicity. The corresponding reactive boundary is described by $\varphi=2 x^4 + 2 (y+1)^4 - 3 x^2 (y+1)^2 - 2=0$, which is represented as the solid green curve in Fig. \ref{fig:swtichvf}. This rather unusual boundary is used to illustrate the generality of our approach to obstacles of various shapes. The desired path is an ellipse described by $\phi = x^2/9+y^2-1=0$. The constant $c$ is $-1.5$ for the repulsive boundary in \eqref{repulsive}, and the gains for the vector fields are $k_p = 1$ and $k_r=0.4$ in \eqref{eqpvf} and \eqref{eqrvf} respectively. One can numerically calculate three equilibria in $\set{E}$: two saddle points and one stable equilibrium. Therefore, a trajectory of the composite vector field will be attracted to the stable equilibria and thus get stuck in the mixed area. However, the switching vector field in \eqref{eqswitch} can resolve this issue. The set $\set{E}^\epsilon$ is approximated using the level set value; i.e., $\set{E}^\epsilon \approx \{ \xi \in \mbr[2]: |\varphi(\xi) - l_2 c / (l_1+l_2)  | \le \epsilon\}$, where $l_1=l_2=\epsilon=0.1$. See Fig. \ref{fig:swtichvf} for the simulation results. %

\subsection{Moving obstacles and Dubins car model in 2D}
\begin{figure}[tb]
	\centering
	\begin{tikzpicture}
	\node[anchor=south west,inner sep=0,xshift=-1.05cm,yshift=-0.75cm] at (0,0) {\includegraphics[width=0.49\columnwidth]{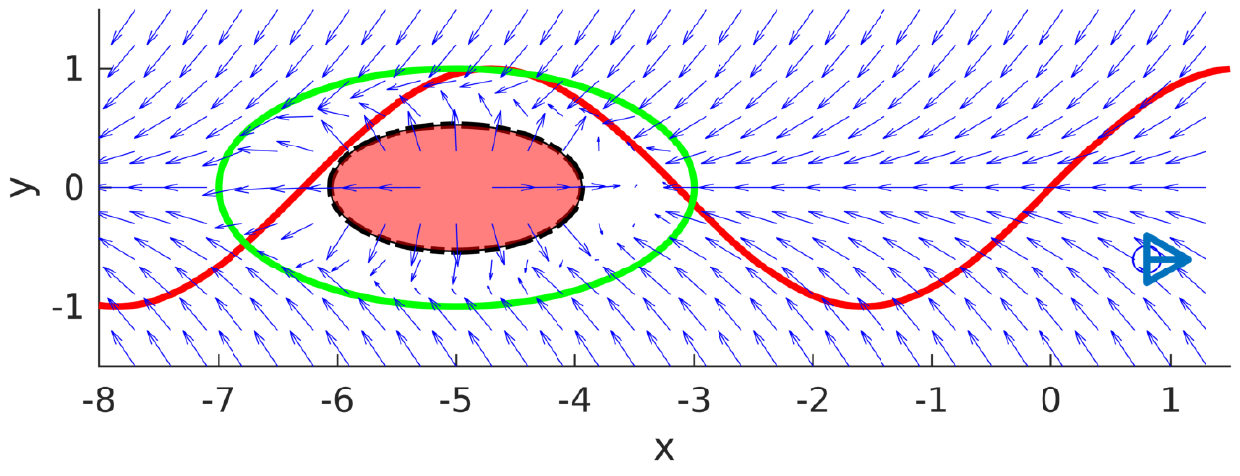}};
	\end{tikzpicture}
	\begin{tikzpicture}
	\node[anchor=south west,inner sep=0,xshift=-1.05cm,yshift=-0.75cm] at (0,0) {\includegraphics[width=0.49\columnwidth]{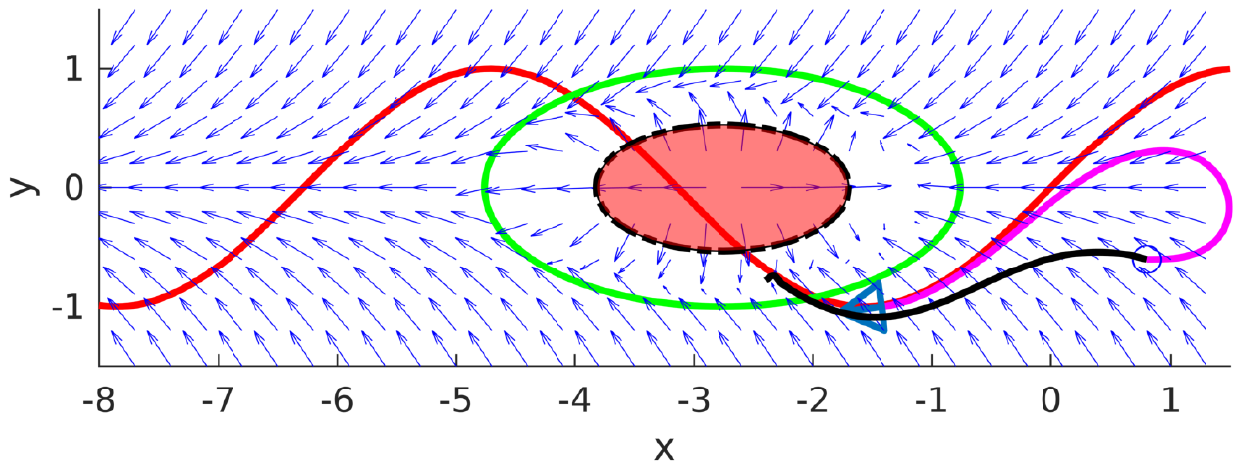}};			
	\end{tikzpicture}
	\begin{tikzpicture}
	\node[anchor=south west,inner sep=0,xshift=-1.05cm,yshift=-0.75cm] at (0,0) {\includegraphics[width=0.49\columnwidth]{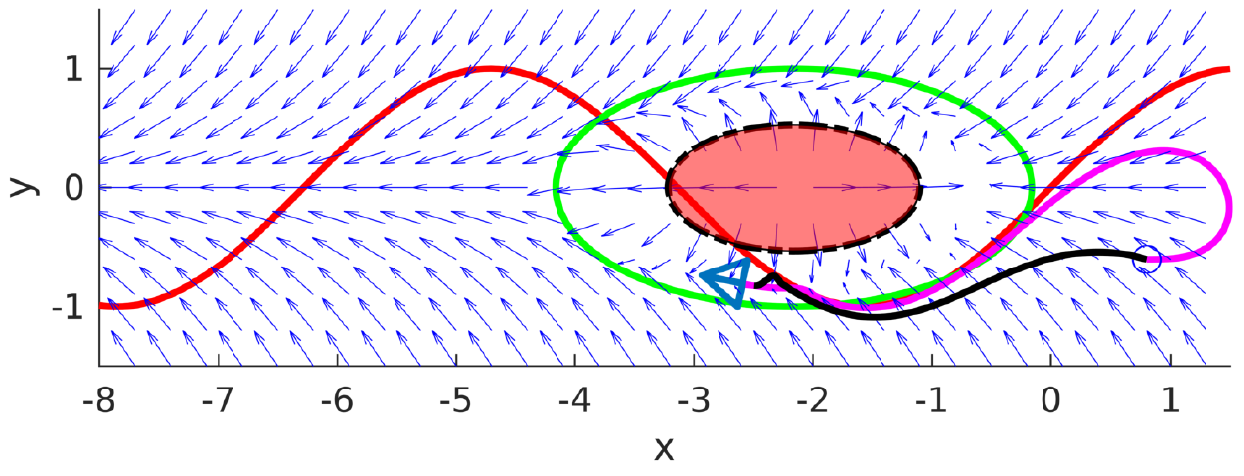}};			
	\end{tikzpicture}
	\begin{tikzpicture}
	\node[anchor=south west,inner sep=0,xshift=-1.05cm,yshift=-0.75cm] at (0,0) {\includegraphics[width=0.49\columnwidth]{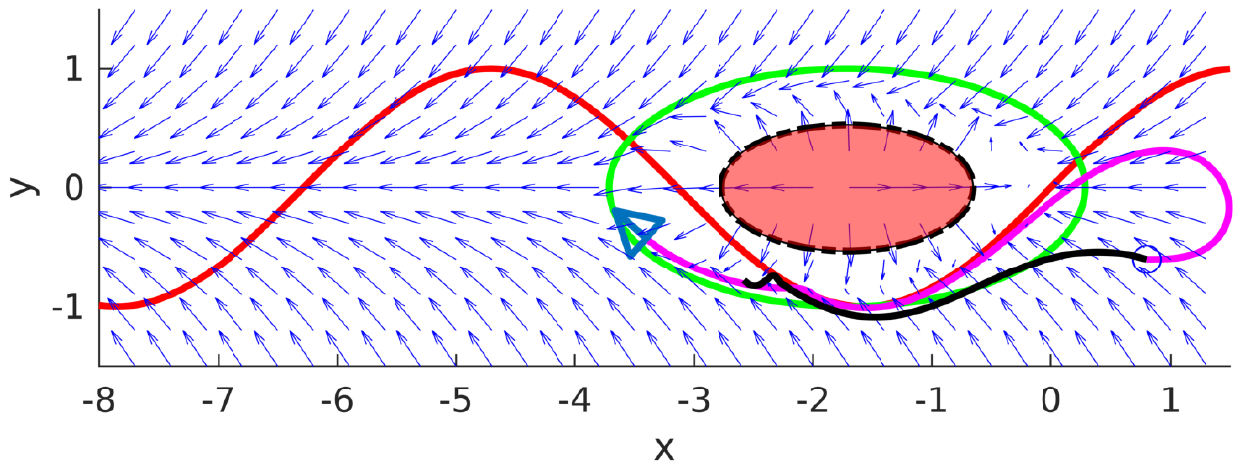}};			
	\end{tikzpicture}
	\begin{tikzpicture}
	\node[anchor=south west,inner sep=0,xshift=-1.05cm,yshift=-0.75cm] at (0,0) {\includegraphics[width=0.49\columnwidth]{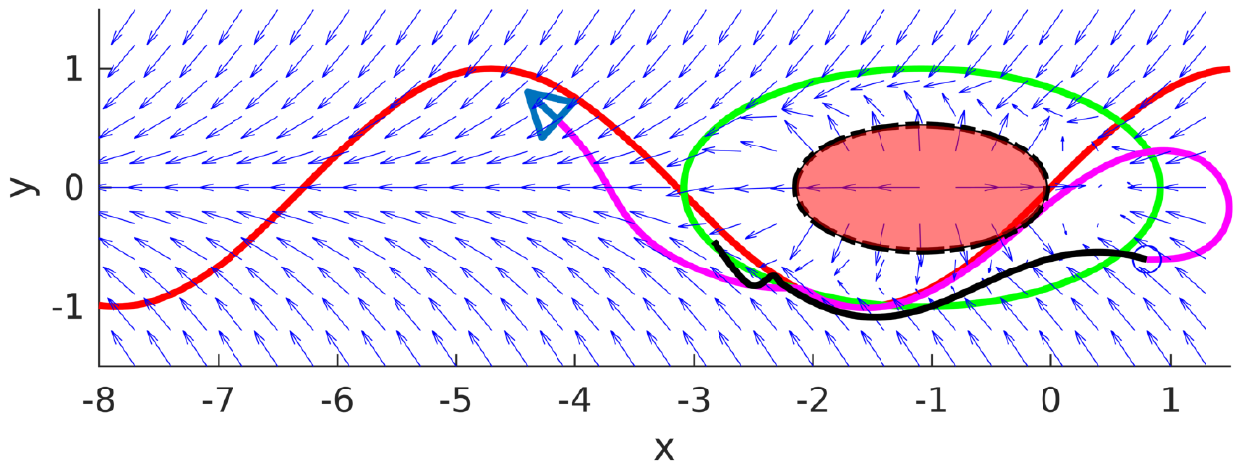}};			
	\end{tikzpicture}
	\begin{tikzpicture}
	\node[anchor=south west,inner sep=0,xshift=-1.05cm,yshift=-0.75cm] at (0,0) {\includegraphics[width=0.49\columnwidth]{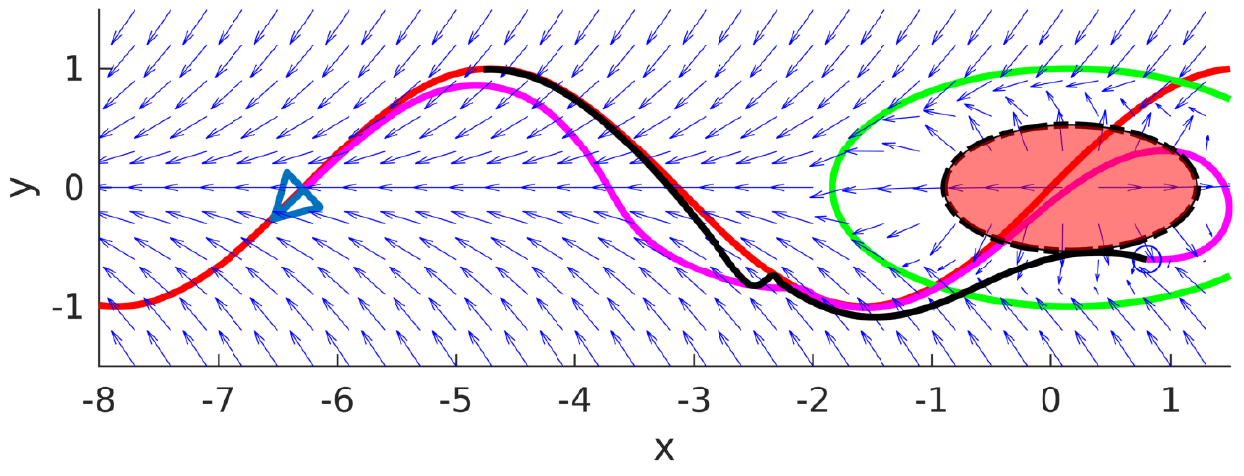}};			
	\end{tikzpicture}
	\caption{Simulation III with a moving obstacle at $t=0$s, $4.47$s,  $5.67$s, $6.57$s, $7.82$s and $10.32$s respectively from left to right, top to bottom. The red curve is the desired path, and the red elliptic shape is the moving obstacle. The repulsive boundary and the reactive boundary are represented by the dashed black line and the solid green line, respectively. The black curve is the trajectory of the single-integrator model starting from $(0.8, -0.6)$, whereas the magenta curve is the trajectory of the 2D Dubins car model with speed $s=1$ and initial conditions $(x_0, y_0, \theta_0)=(0.8, -0.6, 0)$. The triangles represent the poses of the robot at different time instants. The results show that the two trajectories bypass the obstacle and then follow the desired path. }
	\label{fig:moving}
\end{figure}

In this simulation, an obstacle is moving at a constant speed, and thereby the reactive boundary is a moving ellipse. The function $\varphi(t,x,y)$ is $\varphi = (x+5 - v_{\rm obs} t)^2 / a^2 + y^2/b^2 -1$, where $a=2$, $b=1$, and $v_{\rm obs}=0.5$ is the constant speed of the obstacle along the positive x-axis. The desired path is a sinusoidal curve described by $\phi(x,y)=y-\sin(x)=0$. Using the composite vector field, where the reactive vector field is given by \eqref{eqnewvf}, trajectories can bypass the moving obstacle and follow the desired path afterward. To illustrate at the same time the applicability of the composite vector field for robot models other than the single-integrator model, we consider the following 2D Dubins car model: $\dot{x} = s \cos \theta$, $\dot{y} = s \sin \theta$, $\dot{\theta} = u_\theta$, where $(x,y)$ is the robot's position in $\mbr[2]$, $\theta \in [0, 2 \pi)$ is the robot's orientation, $u_\theta$ is the control input and $s$ is the constant speed. We follow \cite[Theorem 4]{yao2020auto} to design the control input $u_\theta$. The simulation results in  Fig. \ref{fig:moving} demonstrate that the proposed algorithm is effective. However, it is still challenging and left for our future work to conduct mathematical analysis on this case where moving obstacles and nonholonomic models are considered altogether.

\subsection{Static obstacles and Dubins car model in 3D}
One of the advantages of using the composite vector field is its natural extension to any higher-dimensional spaces \cite{yao2020singularity}. We take the 3D case as an example. In $\mbr[3]$, a desired path $\set{P}$ is usually described by the intersection of two surfaces defined by the zero-level sets of $\phi_1$ and $\phi_2$ respectively, where $\phi_1,\phi_2: \mbr[3] \to \mbr[]$ are twice continuously differentiable. Namely, the description in \eqref{path} is extended to $\set{P} = \{ \xi \in \mbr[3]: \phi_1(\xi)=0, \phi_2(\xi)=0 \}$, and the path-following vector field becomes
\begin{equation} \label{eq3dpvf}
	\pvf(\xi) = \nabla \phi_1(\xi) \times \nabla \phi_2(\xi) - \sum_{i=1}^{2} k_i \phi_i(\xi) \nabla \phi_i(\xi),
\end{equation}
where $k_i$ are positive constants. As for obstacles in 3D, the corresponding repulsive and reactive boundaries are naturally extended to be \emph{repulsive surfaces} and \emph{reactive surfaces} to avoid collision from all directions in 3D. Similarly, these two surfaces can be defined respectively as the $c$-level surface and zero-level surface of a $C^2$ function $\varphi: \mbr[3] \to \mbr[]$. Similar to \eqref{eq3dpvf}, the reactive vector field $\rvf$ can be defined as
\begin{equation} \label{eq3drvf}
	\rvf(\xi) = \nabla \varphi(\xi) \times v - k_{r} \varphi(\xi) \nabla \varphi(\xi),
\end{equation}
where $v \in \mbr[3]$ is a constant vector indicating how to bypass the obstacle on the reactive surface $\varphi^{-1}(0)$. The 3D composite vector field is attained by substituting $\pvf$ and $\rvf$ in \eqref{eqcompvf} by their 3D counterparts in \eqref{eq3dpvf} and \eqref{eq3drvf} respectively. \cmnt{The extension to higher-dimensional spaces $\mbr[n]$, where $n >3$, is straightforward by using higher-dimensional vector fields in \cite{yao2020singularity}.} However, although the physical intuition is clear, the rigorous analysis in higher-dimensional spaces using dynamical systems theory is challenging; especially, the Poinc{\' a}re-Bendixson theorem is no longer applicable. Exploring the provision of rigorous underpinnings is left for future work, while we provide a simulation example in $\mbr[3]$ below. 

\begin{figure}
	\centering
	\subfigure[3D view]{
		\includegraphics[width=0.48\columnwidth]{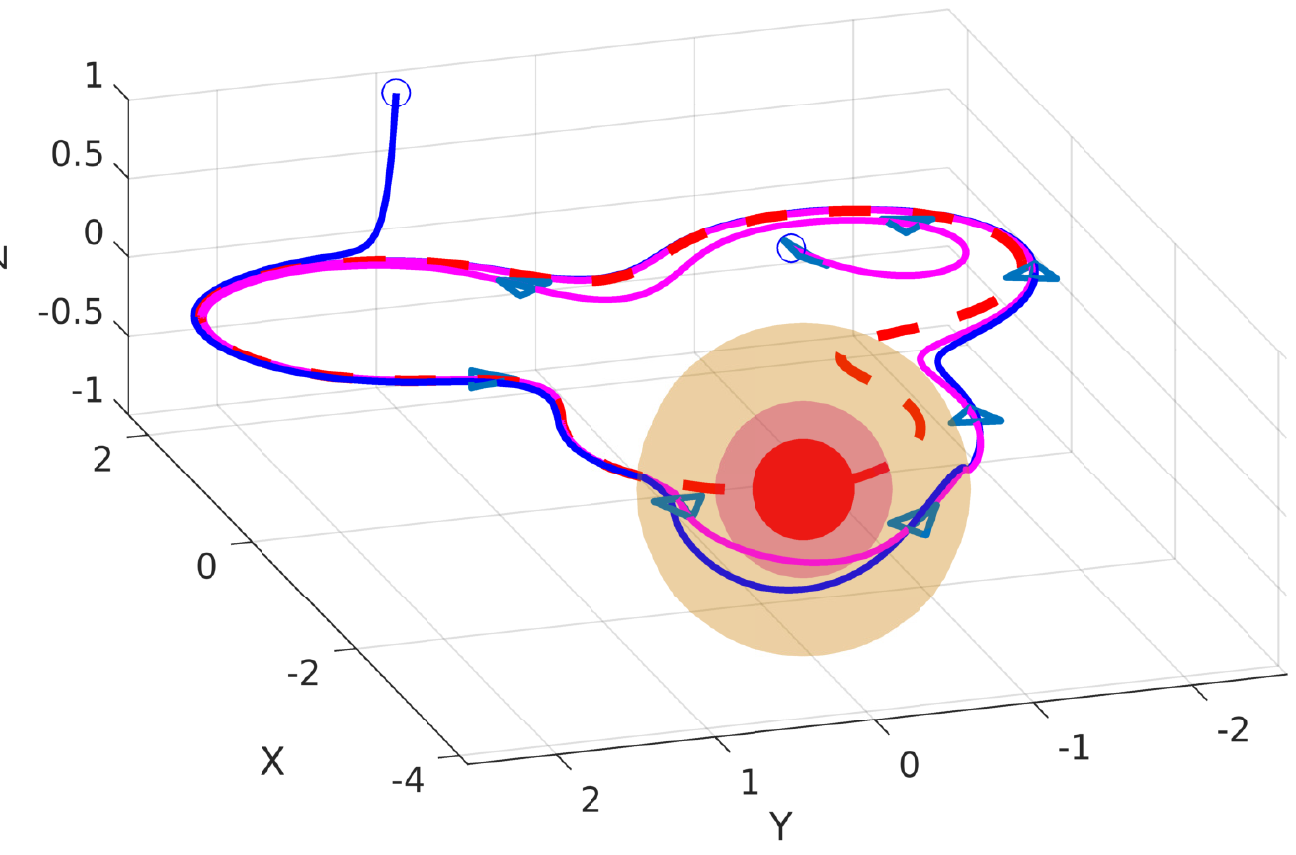}
	}%
	\subfigure[X-Y view]{
		\includegraphics[width=0.48\columnwidth]{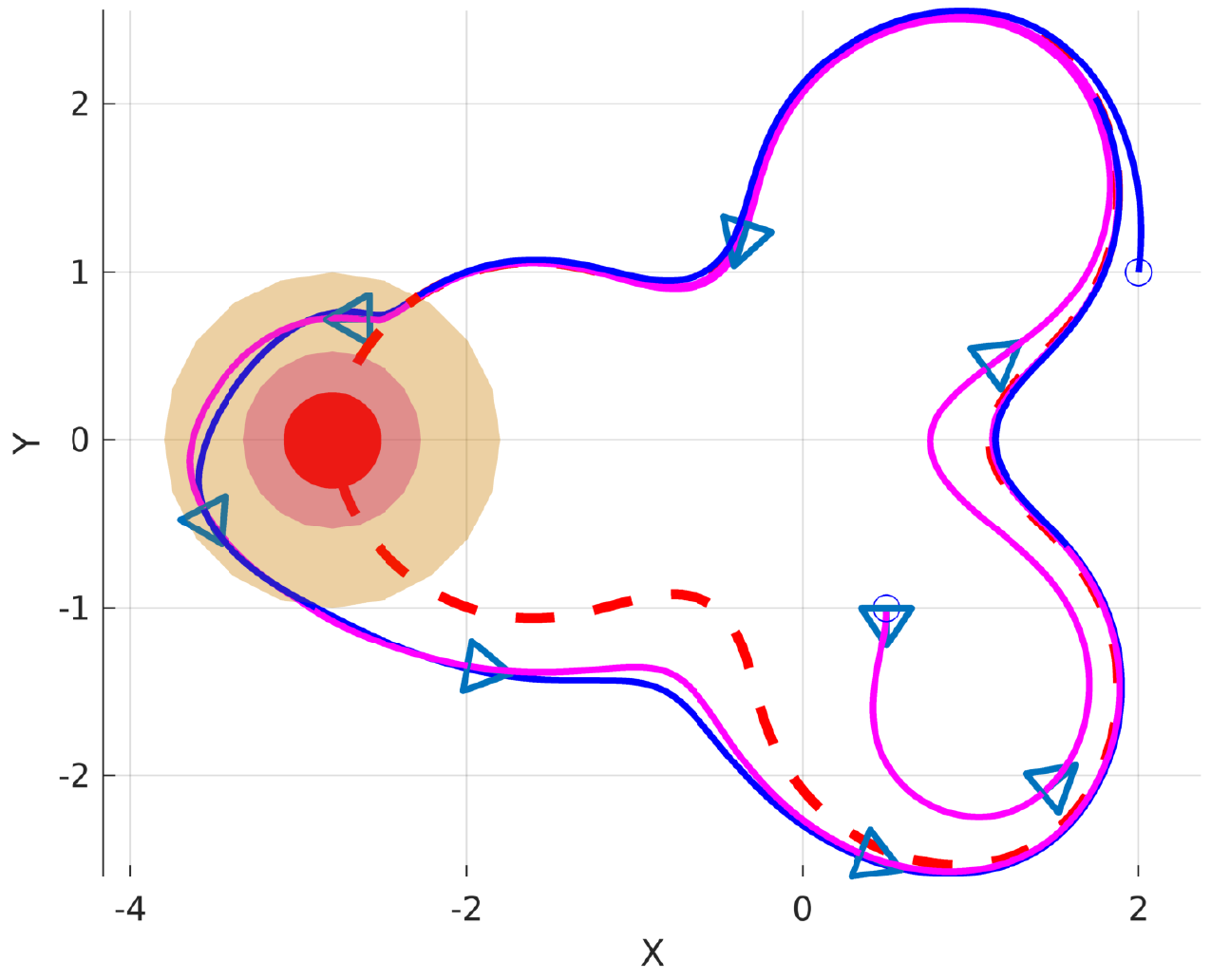}
	}
	\caption{Simulation IV. The solid red ball is the obstacle, and the magenta and orange surfaces are the \emph{repulsive surface} and the \emph{reactive surface} respectively. The blue curve is the trajectory starting at $(2,1,1)$ corresponding to the single-integrator model, whereas the magenta curve is the trajectory with the initial condition $(x_0, y_0, z_0, \theta_0)=(0.5, -1, 0.3, -\pi/2)$ corresponding to the 3D Dubins car model, where the speed is $s=1$. The triangles represent the poses of the robot at different time instants. }
	\label{fig:3dsim}
\end{figure}

In this simulation, a 3D static obstacle modeled by a solid ball occupies a planar desired path shown in Fig. \ref{fig:3dsim}. Since the desired path is a planar curve, we simply choose  $\phi_1(x,y,z)=z$, and  $\phi_2(x,y,z)$ is obtained by using radial-basis functions to interpolate several sample points \cite{Goncalves2010}. Specifically, we assume that $\phi_2(q) = -1 + \sum_{k=1}^{N} \omega_k f(\|q - q^{(k)}\|)$, where $q=(x,y,z) \in \mbr[3]$ is the function input, $q^{(k)} \in \mbr[3]$ are $N$ sample points on the desired path, where $N>1$, $\omega_k$ are $N$ unknown parameters to be calculated and $f: \mbr[]_{\ge 0} \to \mbr[]$ is a radial-basis function. The parameters $\omega_k$ are determined by $N$ constraints: $\phi_2(q^{(k)})=0$ for $k=1,\dots,N$. If the sample points $q^{(k)}$ are chosen such that the Gram matrix $G=[g_{ij}] \in \mbr[N \times N]$, where $g_{ij}=f(\|q^{(i)} - q^{(j)}\|)$, is non-singular, then the parameters are uniquely determined by $\omega = \inv{G} \bm{1}_N$, where $\omega=(\omega_1, \dots, \omega_N)^\top$ and $\bm{1}_N \in \mbr[N]$ is a column vector with all ones. In this example, we choose the radial-basis function $f(r)=r^2 \ln (r+1)$ and six sample points $q^{(k)}$: $(1.5, 0), (1.5, 2.6)$, $(-0.75, 1.3)$, $(-3,0)$,$(-0.75, -1.3)$, $(1.5,-2.6)$. The corresponding parameter vector is $w=(-0.048, 0.035, -0.048, 0.035,  -0.048, 0.035)$. The resulting 3D path-following vector field $\pvf$ is calculated from \eqref{eq3dpvf}. For the reactive surface, we choose $\varphi(x,y,z)=(x+2.8)^2 + y^2 +z^2 - 1$ and  $v=\transpose{(1,0,0)}$ to create the 3D reactive vector field in \eqref{eq3drvf}. The gains for the vector fields are $k_1=5, k_2=2$ and $k_r=2$. We choose $c=-0.72$ for the repulsive surface; i.e., $\repulbdr=\varphi^{-1}(c)$. We choose $l_1=l_2=0.1$ for the zero-in and zero-out bump functions.  %
We also consider the following 3D Dubins car model: $\dot{x} = s \cos \theta$, $\dot{y} = s \sin \theta$, $\dot{\theta} = u_\theta$,  $\dot{z} = u_z$,
where $(x,y,z)$ is the robot's position in $\mbr[3]$, $\theta \in [0, 2 \pi)$ is the robot's orientation, $u_\theta$ and $u_z$ are the control inputs, and $s$ is the constant speed. We follow \cite[Theorem 4]{yao2020auto} to design the control inputs $u_\theta$ and $u_z$. See Fig. \ref{fig:3dsim} for the results.

\section{Conclusion and Future Work} \label{sec_con}
We consider the problem, initially in $\mathbb R^2$, of following an arbitrary desired path occluded by a finite number of static or moving obstacles of arbitrary shapes. This problem, called the VF-CAPF problem, is different from the traditional motion planning problem with obstacles because no starting point and destination point are necessary to calculate a feasible path. We design a composite vector field to solve this problem. To address the issue of motions being ``trapped'' at a stable equilibrium, we propose a switching mechanism involving two vector fields. The path-following and obstacle-avoidance behaviors are provably guaranteed to be effective. 

Our approach using the composite vector field or switching vector field has many advantages; e.g., a) Rigorous theoretical guarantees are provided, which are usually absent in the literature; b) The collision-avoidance behavior is reactive, since the trajectory is guided directly by the vector field, and thereby the operations, such as path re-planning or the creation of a global map, are not required. The vector field is updated easily by adding a new term once a new obstacle is encountered. c) The shapes of the desired path and the obstacles are very general; e.g., convexity and no specific geometric relations are required. d) The composite vector field can be naturally extended to spaces of arbitrary dimensions.

However, there are still some limitations of our algorithm and open problems left for future work. 1) The switching vector field possibly renders trajectories tortuous and complicated, and thus an optimal switching mechanism might be considered.  2) The theoretical analysis regarding the high-dimensional case and nonholonomic models is still an open problem. Some results will be more conservative; e.g., a nonholonomic model may result in the initial increase of the Lyapunov function, which may compromise the initially established safety guarantees for a single-integrator model. 3) To merge the theory with practice, more practical aspects should be considered. For example, \emph{robot-state-dependent} safety zones around robots should be designed for safety laser scanners. It is also crucial to take into account kinematic constraints on the vehicle, such as speed and turning rate limits.

\bibliographystyle{IEEEtran}
\bibliography{ref}

\begin{IEEEbiography}[{\includegraphics[width=1in,height=1.25in,clip,keepaspectratio]{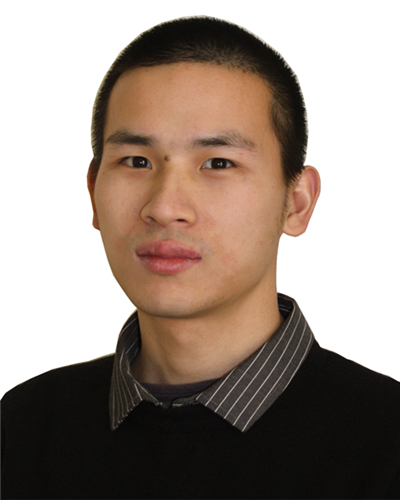}}]{Weijia Yao} obtained the Ph.D. degree with the distinction \textit{cum laude} in systems and control theory from the University of Groningen, Groningen, the Netherlands.
	
His research interests include nonlinear systems and control, robotics and multiagent systems.
	
Dr. Yao was a finalist for the Best Conference Paper Award at ICRA in 2021, a finalist for the Georges Giralt PhD Award in 2022, and the recipient of the Outstanding Master Degree Dissertation award of Hunan province,	China, in 2020.
\end{IEEEbiography}

\begin{IEEEbiography}[{\includegraphics[width=1in,height=1.25in,clip,keepaspectratio]{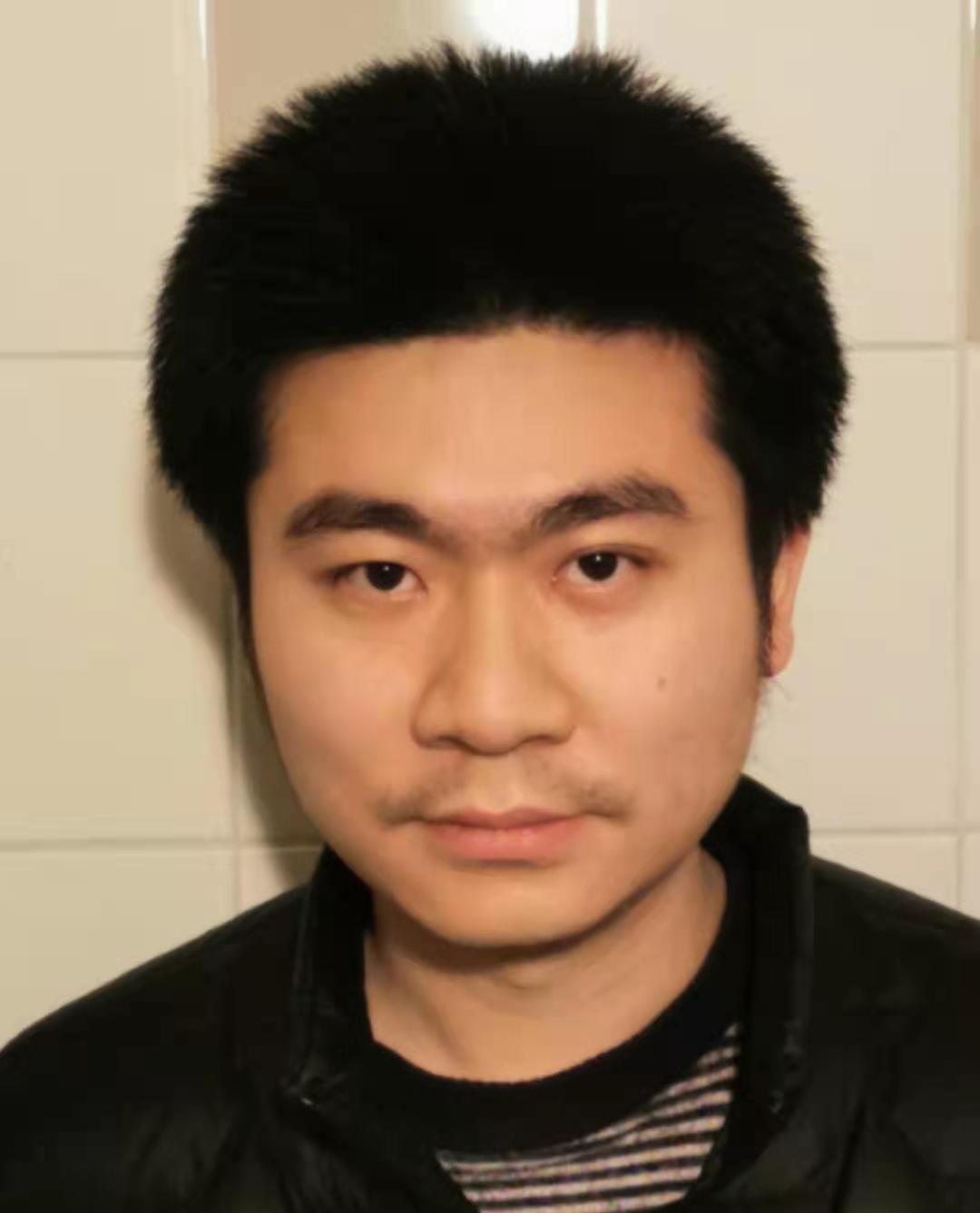}}]{Bohuan Lin} received the bachelor's degree from Sun Yat-sen University, Guangzhou, China, for his first major in theoretical and applied mechanics in 2013, and completed a second major in mathematics in 2014. He then completed the master's degree in mathematics at the School of Mathematics of the same university in 2016 with his work on a certain PDE. He is currently working toward the Ph.D. degree in integrable systems (Hamiltonian/non-Hamiltonian) with the Bernoulli Institute of the University of Groningen, Groningen, the Netherlands.
\end{IEEEbiography}

\begin{IEEEbiography}[{\includegraphics[width=1in,height=1.25in,clip,keepaspectratio]{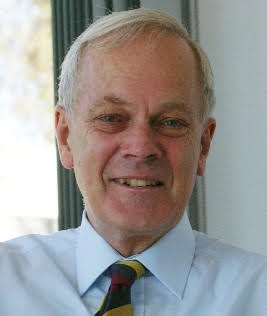}}]{Brian D. O. Anderson} (M’66-SM’74-F’75-LF’07) was born in Sydney, Australia, and educated at Sydney University in mathematics and electrical engineering, with PhD in electrical engineering from Stanford University in 1966.   He is an Emeritus Professor at the Australian National University (having retired as Distinguished Professor in 2016), Distinguished Professor at Hangzhou Dianzi University,  and Distinguished Researcher in Data-61 CSIRO, Australia. His awards include the IEEE Control Systems Award of 1997, the 2001 IEEE James H Mulligan, Jr Education Medal, and the Bode Prize of the IEEE Control System Society in 1992, as well as several IEEE and other best paper prizes. He is a Fellow of the Australian Academy of Science, the Australian Academy of Technological Sciences and Engineering, the Royal Society, and a foreign member of the US National Academy of Engineering. He holds honorary doctorates from a number of universities, including Université Catholique de Louvain, Belgium, and ETH, Zürich. He is a past president of the International Federation of Automatic Control and the Australian Academy of Science. His current research interests are in distributed control, localization and social networks.
\end{IEEEbiography}

\begin{IEEEbiography}[{\includegraphics[width=1in,height=1.25in,clip,keepaspectratio]{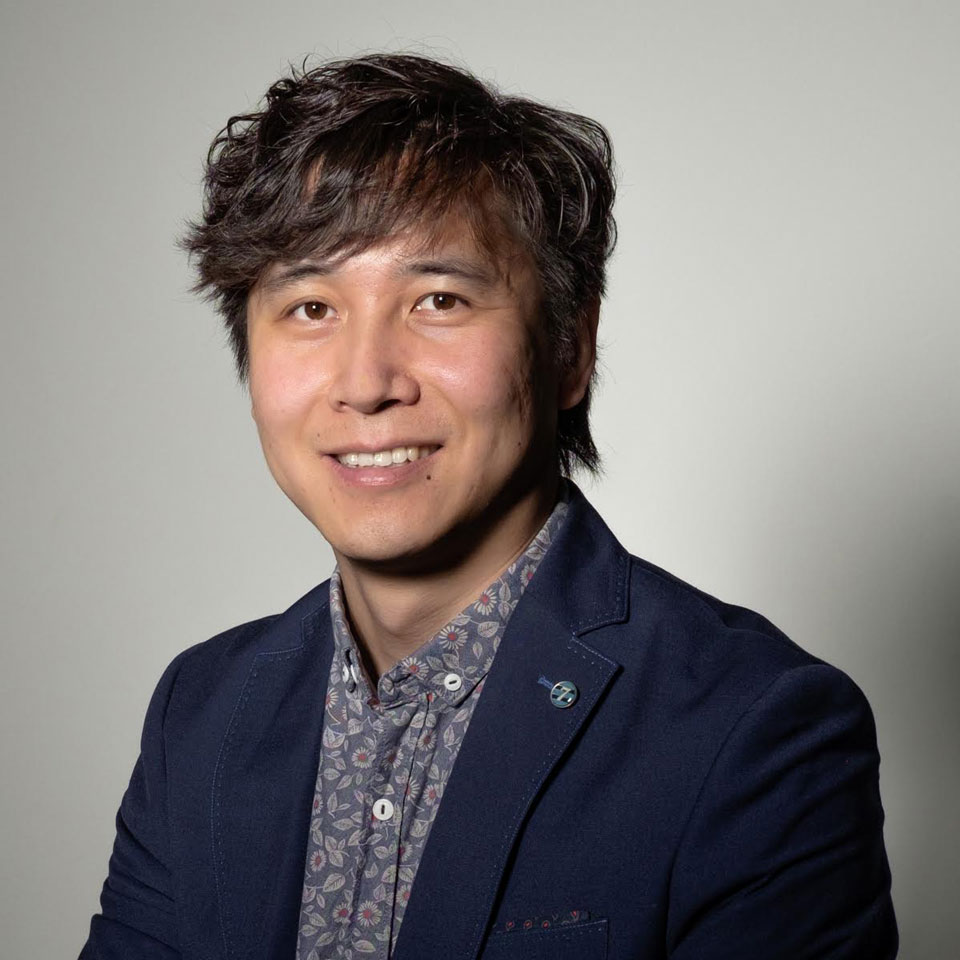}}]{Ming Cao}  has since 2016 been a professor of systems and control with the Engineering and Technology Institute (ENTEG) at the University of Groningen, the Netherlands, where he started as a tenure-track Assistant Professor in 2008. He received the Bachelor degree in 1999 and the Master degree in 2002 from Tsinghua University, Beijing, China, and the Ph.D. degree in 2007 from Yale University, New Haven, CT, USA, all in Electrical Engineering. From September 2007 to August 2008, he was a Postdoctoral Research Associate with the Department of Mechanical and Aerospace Engineering at Princeton University, Princeton, NJ, USA. He worked as a research intern during the summer of 2006 with the Mathematical Sciences Department at the IBM T. J. Watson Research Center, NY, USA. He is the 2017 and inaugural recipient of the Manfred Thoma medal from the International Federation of Automatic Control (IFAC) and the 2016 recipient of the European Control Award sponsored by the European Control Association (EUCA). He is a Senior Editor for Systems and Control Letters, and an Associate Editor for IEEE Transactions on Automatic Control, IEEE Transactions on Circuits and Systems and IEEE Circuits and Systems Magazine. He is a vice chair of the IFAC Technical Committee on Large-Scale Complex Systems. His research interests include autonomous agents and multi-agent systems, complex networks and decision-making processes.   
\end{IEEEbiography}
\vfill

\end{document}